\documentclass[superscriptaddress,twocolumn]{revtex4-2}

\usepackage{graphicx}
\usepackage{amsmath}
\usepackage{braket}
\usepackage{array}
\usepackage{color}

\newcommand{\localedge}{v}
\newcommand{\localedgevector}{\mathbf{\localedge}}
\newcommand{\localedgedirection}{\hat{\localedge}}
\newcommand{\globaledge}{r}
\newcommand{\globaledgevector}{\mathbf{\globaledge}}
\newcommand{\globaledgedirection}{\hat{\globaledge}}
\newcommand{\permutation}{\mathcal{P}}
\newcommand{\latticevector}{\boldsymbol{\ell}}
\newcommand{\sectorangle}{\alpha}
\newcommand{\angularvelocity}{\boldsymbol{\omega}}
\newcommand{\vertexfold}{\phi}
\newcommand{\facebend}{\tau}
\newcommand{\vertexdisplacement}{\mathbf{u}}
\newcommand{\foldcoefficient}{\zeta}
\newcommand{\vertexamplitude}{\mathcal{V}}
\newcommand{\faceamplitude}{\mathcal{F}}
\newcommand{\compatibility}{\mathbf{C}}
\newcommand{\globalcoefficient}{\chi}
\newcommand{\edgeproduct}{R}
\newcommand{\latticevectorchange}{\boldsymbol{\Delta}}
\newcommand{\latticestrain}{\epsilon}
\newcommand{\latticecurvature}{\kappa}
\newcommand{\latticerotation}{\boldsymbol{\Omega}}
\newcommand{\normalvector}{\mathbf{N}}
\newcommand{\poisson}{\nu}
\newcommand{\dihedralangle}{\gamma}
\newcommand{\spherenorm}[1]{\lVert{#1}\rVert}

\begin{document}

\title{Discrete symmetries control mechanical response in parallelogram-based origami}
\author{James McInerney}
\affiliation{School of Physics, Georgia Institute of Technology, Atlanta, GA 30332}
\affiliation{Department of Physics, University of Michigan, Ann Arbor, MI 48109} 
\author{Glaucio H. Paulino}
\affiliation{School of Civil Engineering, Georgia Institute of Technology, Atlanta, GA 30332} 
\affiliation{Department of Civil and Environmental Engineering, Princeton University, Princeton, NJ, 08544} 
\affiliation{Princeton Institute for the Science and Technology of Materials, Princeton University, Princeton, NJ, 08544} 
\affiliation{Department of Mechanical and Aerospace Engineering, Princeton University, Princeton, NJ, 08544} 
\author{D. Zeb Rocklin}
\affiliation{School of Physics, Georgia Institute of Technology, Atlanta, GA 30332} 
\date{\today}
\begin{abstract}
Geometric compatibility constraints dictate the mechanical response of soft systems that can be utilized for the design of mechanical metamaterials such as the negative Poisson ratio Miura-ori origami crease pattern. Here, we develop a formalism for linear compatibility that enables explicit investigation of the interplay between geometric symmetries and functionality in origami crease patterns. We apply this formalism to a particular class of periodic crease patterns with unit cells composed of four arbitrary parallelogram faces and establish that their mechanical response is characterized by an anticommuting symmetry. In particular, we show that the modes are eigenstates of this symmetry operator and that these modes are simultaneously diagonalizable with the symmetric strain operator and the antisymmetric curvature operator. This feature reveals that the anticommuting symmetry defines an equivalence class of crease pattern geometries which possess equal and opposite in-plane and out-of-plane Poisson's ratios.
\end{abstract}
\maketitle

Discrete symmetries characterize the properties of physical systems~\cite{weyl1950theory,dresselhaus2007group} ranging from topological insulators and superconductors~\cite{altland1997nonstandard, schnyder2008classification, kitaev2009periodic, ryu2010topological} to frustrated magnets~\cite{roychowdhury2018classification} and mechanical metamaterials~\cite{susstrunk2016classification}. Such symmetries can dictate the rigid deformations of periodic mechanical networks~\cite{guest2014}, including the generation of negative Poisson's ratio (auxetic) modes~\cite{mitschke2013symmetry,fowler2014symmetry}. Such auxetic mechanical metamaterials are characterized by either dilational strain or synclastic (dome-like) curvature, and are desirable for a variety of applications in engineering that cannot be achieved with conventional materials~\cite{babaee20133d,hou2015metamaterials,kolken2017auxetic,kelkar2020cellular}. While this dilational in-plane behavior of auxetic materials typically couples to the synclastic out-of-plane behavior~\cite{landau1959course}, a family of highly-symmetric origami crease patterns exhibits equal and \emph{opposite} in-plane and out-of-plane Poisson's ratios~\cite{wei2013geometric,schenk2013geometry,nassar2017curvature,pratapa2019geometric}.

Origami-inspired structures are thin sheets whose low-energy mechanical response is dictated by the geometry of their crease patterns~\cite{peraza2014origami,rogers2016origami,santangelo2017extreme,santangelo2020theory} via compatibility conditions that restrict deformations to \emph{isometries} which transform the sheets without stretching their faces~\cite{huffman1976curvature,hull2002modelling}. Modern fabrication and actuation techniques can be used to self fold such structures at the macro-scale~\cite{hawkes2010programmable,tolley2014self,melancon2021multistable} as well as the nano- and micro-scales~\cite{bassik2009microassembly,cho2011nanoscale,lazarus2019self,liu2012self,na2015programming,liu20162d,liu2017sequential,lin2020folding,miskin2018graphene} for applications in engineering~\cite{turner2016review,meloni2021engineering} and soft robotics~\cite{kim2018origami,kim2019bioinspired,novelino2020untethered,wu2021stretchable}. These isometric configurations can be considered as degenerate ground states  that rely on symmetries of the crease pattern to rigidly fold~\cite{tachi2009generalization,waitukaitis2015origami,evans2015rigidly,feng2020designs,dieleman2020jigsaw}; however, interplay between these symmetries and the the functionality of origami as mechanical metamaterials~\cite{lv2014origami} with negative Poisson's ratios~\cite{wei2013geometric,schenk2013geometry,nassar2017curvature,pratapa2019geometric} or high stiffness-to-weight ratios~\cite{filipov2015origami} has not been explored explicitly. 

The linear isometries of an origami sheet are conventionally modeled via the rigid deformations of its triangulation~\cite{schenk2010origami,filipov2017bar}, where ``virtual'' creases are introduced to accommodate bending of the faces. Such rigid deformations can be specified either via vertex displacements that do not stretch any physical or virtual creases~\cite{evans2015lattice} (as commonly applied to mechanical networks composed of point masses and central-force springs) or via folding about the creases that does not rotate elements of the sheet relative to themselves~\cite{gluck1975almost,crapo1982statics}. This latter approach has been applied to the Morph family of parallelogram-based crease patterns~\cite{pratapa2019geometric}, expanding upon previous investigations of the Miura-ori~\cite{schenk2013geometry,wei2013geometric} and eggbox~\cite{nassar2017curvature} crease patterns, thereby showing both analytically and numerically the existence of equal and opposite in-plane and out-of-plane Poisson's ratios. While it is straightforward to show numerically that this result holds for the crease patterns with more generic parallelogram faces discussed herein, analytical solutions become intractable in the absence of multiple symmetries that obfuscate the underlying physical principle.

In this work, we introduce an alternative model for the linear isometries of parallelogram-based origami that yields simple analytic formulae for the in-plane and out-of-plane Poisson's ratios and elucidates their equal and opposite relationship, for the unexplored periodic crease patterns with unit cells composed of four arbitrary parallelogram faces. We first introduce this family of four-parallelogram origami that generalizes the Miura-ori~\cite{schenk2013geometry,wei2013geometric}, the eggbox~\cite{nassar2017curvature}, and the Morph crease patterns~\cite{pratapa2019geometric}. We then derive compatibility constraints for the linear isometries and show that they possess an anticommuting symmetry which constrains the linear isometries at both the \emph{intra}cellular and \emph{inter}cellular scales. This leads us to our key result that the system has a symmetric linear isometry with an in-plane Poisson's ratio which is equal and opposite to the out-of-plane Poisson's ratio for the system's antisymmetric linear isometry, implying that one is always negative. Finally, we explore how these Poisson's ratios may change sign as the crease pattern rigidly folds along its one-dimensional configuration space of degenerate ground states. We close with concluding remarks that address extension to future work and experimental implementation.

\begin{figure*}
    \centering
    \includegraphics{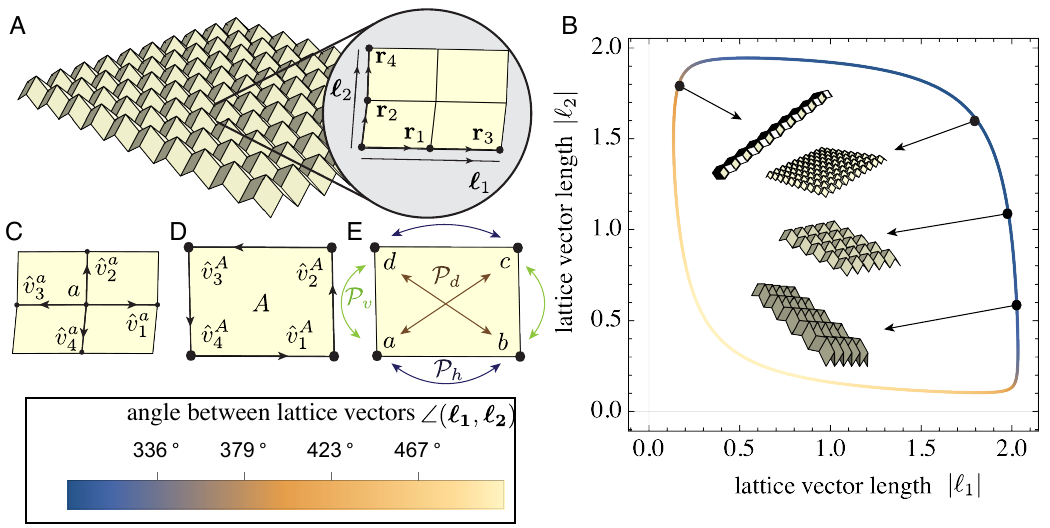}
    \caption{(A) An example four-parallelogram origami tessellation and its unit cell characterized by the four edge vectors, $\globaledgevector_1, \globaledgevector_2, \globaledgevector_3$, and $\globaledgevector_4$, that specify the two generically non-orthogonal lattice vectors, $\latticevector_1 = \globaledgevector_1 + \globaledgevector_3$ and $\latticevector_2 = \globaledgevector_2 + \globaledgevector_4$. (B) Degenerate ground states of the four-parallel origami shown in panel A where the color indicates the angle between the lattice vectors. Local notation for edges at the (C) vertex and (D) face specified by the superscript where the subscript specifies the local edge index which increases cyclically in counterclockwise order. (E) Labeling of the vertices in the unit cell under various two-cycle permutations, $\permutation_h$, $\permutation_v$, and $\permutation_d$, each satisfying $\permutation^2 = 1$.}
    \label{fig:system}
\end{figure*}

\section*{Results and discussion}

\subsection*{Four-parallelogram origami}

Investigations of origami sheets typically rely on highly symmetric crease patterns, such as the renowned Miura-ori, to simplify the analysis. Here, we introduce the completely generic family of four-parallelogram origami and discuss the sole symmetry which governs its members' linear response, including the special cases of the Miura-ori~\cite{schenk2013geometry,wei2013geometric}, the eggbox~\cite{nassar2017curvature}, and the Morph~\cite{pratapa2019geometric}.

Consider spatially-periodic origami sheets composed of unit cells with four arbitrary parallelogram faces such as that shown in Fig.~\ref{fig:system}A. These unit cells are defined by the four unique edges, $\globaledgevector_i$ and its vertices are generically non-developable which means that they cannot necessarily be folded from a single sheet of material. In contrast to the aforementioned special cases, the lattice vectors, $\latticevector_1 = \globaledgevector_1 + \globaledgevector_3$ and $\latticevector_2 = \globaledgevector_2 + \globaledgevector_4$, are generically non-orthotropic, $\latticevector_1 \cdot \latticevector_2 \neq 0$; moreover, the lattice vectors stretch and shear along the one-dimensional manifold of degenerate ground states (see App.~\ref{appendix:system}) such as that shown in Fig.~\ref{fig:system}B. Since periodic origami composed of arbitrary quadrilateral faces cannot be rigidly folded~\cite{mcinerney2020hidden}, these crease patterns possess a symmetry that renders its constraints redundant~\cite{dieleman2020jigsaw} which can be characterized by introducing local notation for the edges.

Let $\localedgevector_i^a$ denote the edges emanating from vertex $a$ where $i$ cyclically labels the edges in counterclockwise order as shown in Fig.~\ref{fig:system}C. For either of the two pairs of non-adjacent vertices in the unit cell, the same edges that emanate from one enter the other. Consequently, the locally defined edges are antisymmetric under the permutation operation $\permutation_d = \permutation_h \permutation_v = \permutation_h \permutation_v$ introduced in Fig.~\ref{fig:system}E that swaps non-adjacent vertices: $\localedgevector_i^{\permutation_d a} = - \localedgevector_{i+2}^a$. Note that these permutations are two-cycles, $\permutation^2 = 1$, and are isomorphic to the two-fold rotoinversion spatial symmetry of the crease pattern.

Similarly, let $\localedgevector_i^A$ denote the edges bounding a face where $A$ labels the face and $i$ cyclically labels the edge in counterclockwise order as shown in Fig.~\ref{fig:system}D. Since the faces are all parallelograms, every other edge is anti-parallel, $\localedgevector_{i+2}^A = - \localedgevector_i^A$. Consequently, the interior (sector) angles of the face are supplementary at adjacent vertices and identical at non-adjacent vertices. In contrast to the vertex symmetry of the four-parallelogram, this feature holds for crease patterns with arbitrary numbers of paralellogram faces, which has implications for the isometries discussed in the next section.

\subsection*{Linear isometries and compatibility conditions}

The low-energy response of origami sheets is dominated by linear isometries which rotate elements of the sheet without stretching them. Such rotations include both folding at the creases and bending of the faces which are treated equivalently in previous works~\cite{wei2013geometric,schenk2013geometry,nassar2017curvature,pratapa2019geometric}. Here, we derive compatibility conditions that distinguish between these two types of local isometries and show that the linear isometries of parallelogram-based origami are governed by a compatibility matrix that exclusively depends on vertex folding.

\begin{figure}
    \centering
    \includegraphics{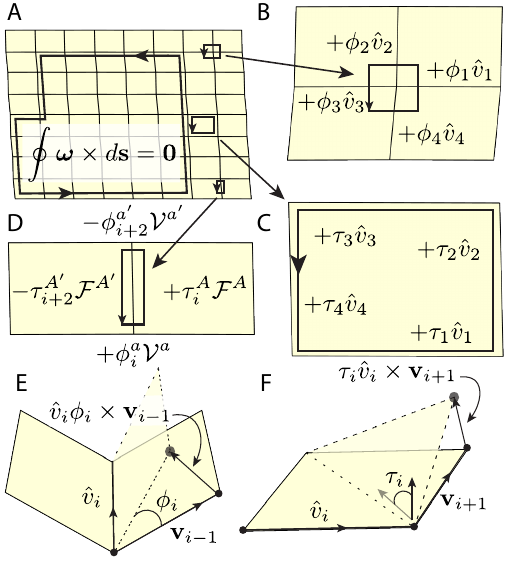}
    \caption{Compatibility conditions for linear isometries. (A) The loop integral of the angular velocity must vanish over arbitrary closed paths. (B) Folding about crease $i$ in the vicinity of vertex $a$ induces a rotation generated by $\vertexfold_i^a \localedgedirection_i^a$ that accumulates along subsequent creases and must vanish over a closed path as dictated by Eq.~(\ref{eq:vertexcompatibility}). (C) Applying a torsion along crease $i$ on face $A$ induces a rotation generated by $\facebend_i^A \localedgedirection_i^A$ as well as a displacement $\facebend_i^A \localedgedirection_i^A \times \localedgevector_{i+1}^A$ that accumulates along subsequent creases and must vanish over a closed path as dictated by Eq.~(\ref{eq:facecompatibility}). (D) The vertex folds and face torsions couple via compatibility in Eq.~(\ref{eq:edgecompatibility}). Schematic of (E) vertex folding and (F) face bending along with the displacements these rotations generate.}
    \label{fig:compatibility}
\end{figure}

The linear isometries are described by a spatially varying, temporally static \emph{angular velocity field}, $\angularvelocity$, that rotates infinitesimal elements of the sheet, $d \mathbf{X}$ relative to one another. Since these rotations must preserve the length of any closed loop on the surface of the sheet, $\mathbf{X} = \mathbf{X}(x_1,x_2)$, (such as that shown in Fig.~\ref{fig:compatibility}A) the angular velocity is constrained by the compatibility condition: 

\begin{equation} \label{eq:closure}
    \oint \angularvelocity \times d \mathbf{X} = \mathbf{0}:
\end{equation} 

\noindent
 Application of Stokes's theorem to Eq.~(\ref{eq:closure}) over a single face shows that the angular velocity must lie in the plane of the face to satisfy $\partial_1 \omega_3 = \partial_2 \omega_3 = 0$ and hence, must point along the edges at the boundary to match with the adjacent faces. Thus, the interior of the face can undergo any linear isometry of a plane provided that the angular velocity is compatible between different corners of the sheet. 

Two corners of the origami sheet, $(a,A)$ and $(a_0,A_0)$, are connected via a sequence of paths \emph{across} and \emph{along} creases. The angular velocity gradient between the corners of two edge-sharing faces, $\angularvelocity^{(a,A')} - \angularvelocity^{(a,A)} = \pm \vertexfold_i^a \localedgedirection_i^a$, corresponds to folding over the crease (see Fig.~\ref{fig:compatibility}E) while similarly, the angular velocity gradient between the corners of two vertices which share both a face and an edge, $\angularvelocity^{(a',A)} - \angularvelocity^{(a,A)} = \pm \facebend_i^A \localedgedirection_i^A$, corresponds to applying a torsion along the crease (see Fig.~\ref{fig:compatibility}F). By convention, the sign is positive (negative) when the crease is traversed along (against) the ordering of the local edges in both cases.

The angular velocity gradients and the displacements that they generate accumulate along the path between the two corners:

\begin{equation} \label{eq:angularvelocity}
    \angularvelocity^{(a,A)} - \angularvelocity^{(a_0,A_0)} = \sum_{i',a'} \pm \vertexfold_{i'}^{a'} \localedgedirection_{i'}^{a'} + \sum_{i', A'} \pm \facebend_{i'}^{A'} \localedgedirection_{i'}^{A'},
\end{equation}
\begin{equation} \label{eq:vertexdisplacement}
    \vertexdisplacement^{(a,A)} - \vertexdisplacement^{(a_0,A_0)} = \sum_{i',a',A'} (\angularvelocity^{(a',A')} - \angularvelocity^{(a_0,A_0)}) \times \localedgevector_{i'}^{a'}. 
\end{equation}

\noindent
Note that since the vertex displacements contain a double summation in Eq.~(\ref{eq:vertexdisplacement}), they grow quadratically between cells, implying that the angular velocity field generically induces some amount of intercellular curvature.  These folds and torsions must satisfy the compatibility condition in Eq.~(\ref{eq:closure}) over any closed sequence of corners. Such a sequence constrains the folds in the vicinity of a vertex, as illustrated in Fig.~\ref{fig:compatibility}B, and the torsions on the boundary of a face, as illustrated in Fig.~\ref{fig:compatibility}C, while the edges couple the folds to the torsions as illustrated in Fig.~\ref{fig:compatibility}D:

\begin{equation} \label{eq:vertexcompatibility}
    \sum_{i'} \vertexfold_{i'}^a \localedgedirection_{i'}^a = \mathbf{0}, 
\end{equation}
\begin{equation} \label{eq:facecompatibility}
    \sum_{i'} \facebend_{i'}^A \localedgedirection_{i'}^A = \mathbf{0}, \quad \facebend_i^A \localedgedirection_i^A \times \localedgevector_{i+1}^A = \facebend_{i-1}^A \localedgedirection_{i-1}^A \times \localedgevector_{i-2}^A,
\end{equation}
\begin{equation} \label{eq:edgecompatibility}
    \vertexfold_i^a + \facebend_j^A - \vertexfold_{i+2}^{a'} - \facebend_{j+2}^{A'} = 0.
\end{equation}

\noindent
Importantly, the vertex and face compatibility conditions in Eqs.~(\ref{eq:vertexcompatibility}) and~(\ref{eq:facecompatibility}) respectively admit one-dimensional analytical solutions. Note that these face compatibility conditions only admit trivial solutions for triangular faces for which the isometries are restricted to vertex folding.

Projecting the cross product of the directions of any two edges emanating from the vertex, $\localedgedirection_i^a \times \localedgedirection_j^a$, onto Eq.~(\ref{eq:vertexcompatibility}) reveals that the vertex folds are proportional to triple products of edge directions. For parallelogram faces, $\localedgevector_i^A = - \localedgevector_{i+2}^A$, the torsions must be proportional to their edge lengths to satisfy Eq.~(\ref{eq:facecompatibility}). These proportionality factors are respectively the vertex amplitudes, $\vertexamplitude^a$, and the face amplitudes, $\faceamplitude^A$, that determine the local solutions:

\begin{equation} \label{eq:vertexsolution}
    \vertexfold_i^a = (-1)^i \vertexamplitude^a \foldcoefficient_i^a, \quad \foldcoefficient_i^a \equiv \localedgedirection_{i+1}^a \cdot \localedgedirection_{i+2}^a \times \localedgedirection_{i+3}^a,
\end{equation}
\begin{equation} \label{eq:facesolution}
    \facebend_i^A  = (-1)^i \faceamplitude^A \localedge_i^A.
\end{equation}

\noindent
Note that while the vertex amplitudes have no units, the face amplitudes must have units of inverse length to ensure that the angular velocity is dimensionless. Substitution of the local solutions from Eqs.~(\ref{eq:vertexsolution},~\ref{eq:facesolution}) into the edge compatibility condition, Eq.~(\ref{eq:edgecompatibility}), yields 

\begin{equation} \label{eq:edgecoupling}
\vertexamplitude^a \foldcoefficient_i^a - \vertexamplitude^{a'} \foldcoefficient_{i+2}^{a'} + ( \faceamplitude^A - \faceamplitude^{A'}) \localedge_i^a = 0,
\end{equation}

\noindent
which admits solutions with uniform face amplitudes and zero vertex amplitudes henceforth referred to as the uniform face-bending mode. Consequently, the appropriate linear combination of the edge compatibility conditions from Eq.~(\ref{eq:edgecompatibility}) on the four subsequent edges around a single vertex eliminates the face amplitudes altogether:

\begin{equation} \label{eq:isometrycondition}
    \sum_{i'} \Big( \frac{\foldcoefficient_{i'}^a}{\localedge_{i'}^a} \vertexamplitude^a - \frac{\foldcoefficient_{i'+2}^{a'}}{\localedge_{i'+2}^{a'}} \vertexamplitude^{a'} \Big) = 0.
\end{equation}

The amplitudes that satisfy Eq.~(\ref{eq:isometrycondition}) at every vertex are linear isometries whose corresponding face amplitudes are determined by inverting the edge compatibility condition in Eq.~(\ref{eq:edgecoupling}) (see App.~\ref{appendix:compatibility}). These isometries are equivalent to those modeled by a triangulation; hence, there are generically two such modes in addition to the uniform face-bending mode~\cite{mcinerney2020hidden}. These contraints can be concatenated to compose the \emph{compatibility} matrix, $\compatibility$, that maps the vertex amplitudes, $\ket{\vertexamplitude}$, to the constraint in Eq.~(\ref{eq:isometrycondition}) at every vertex so that the linear isometries span the matrix's nullspace $\compatibility \ket{\vertexamplitude} = \mathbf{0}$. In this notation, the ``bra'', $\bra{\vertexamplitude_1}$, and ``ket'', $\ket{\vertexamplitude_2}$, have the inner product $\braket{\vertexamplitude_1 | \vertexamplitude_2} = \boldsymbol{\vertexamplitude}_1^\dagger \cdot \boldsymbol{\vertexamplitude}_2$ and transform under matrix operators as $\bra{\vertexamplitude} \compatibility = \bra{ \compatibility \vertexamplitude}$ and $\compatibility \ket{\vertexamplitude} = \ket{ \compatibility \vertexamplitude}$ respectively.

The uniform face-bending mode, $\ket{\vertexamplitude} = \mathbf{0}$, satisfying edge compatibility in Eq.~(\ref{eq:edgecoupling}) with uniform face amplitudes is a nontrivial linear isometry that exists for crease patterns with \emph{any} number of parallelogram faces; hence, the compatibility condition in Eq.~(\ref{eq:isometrycondition}) is also valid for any parallelogram-based origami crease pattern. However, this is not true for generic quadrilateral faces because the face torsions in Eq.~(\ref{eq:facesolution}) no longer satisfy the face compatibility conditions in Eq.~(\ref{eq:facecompatibility}).

\subsection*{Vertex permutation symmetry in the four-parallelogram}

The spatial symmetry of four-parallelogram origami constrains the linear isometries which satisfy the compatibility conditions in Eq.~(\ref{eq:isometrycondition}). While such modes generically vary between cells according to Bloch's theorem~\cite{evans2015lattice,mcinerney2020hidden}, we show here that the two non-trivial, uniform isometries are eigenstates of the permutation operator, $\permutation_d$, with opposite eigenvalues. Our analysis is simplified by our self-adjoint compatibility matrix which anticommutes with with the permutation operator. This feature contrasts with previous work where the compatibility matrix maps between vector spaces that transform differently under symmetries of the sheet.

The antisymmetry of the local edge vectors, $\localedgevector^{\permutation_d a} = - \localedgevector_{i+2}^a$, implies that the local folding coefficients are antisymmetric between nonadjacent vertices, $\foldcoefficient_i^{\permutation_d a} = - \foldcoefficient_{i+2}^a$. Hence, the compatibility matrix constructed from Eq.~(\ref{eq:isometrycondition}) anticommutes with the permutation operator:

\begin{equation} \label{eq:anticommutation}
    \permutation_d \compatibility \permutation_d = - \compatibility.
\end{equation} 

\noindent
The eigenbasis of this permutation is represented by the basis vectors $\ket{\pm \pm}$ where the first (second) sign indicates the eigenvalue of the basis vector under the horizontal (vertical) permutation $\permutation_{h(v)}$ illustrated in Fig.~\ref{fig:system}E:

\begin{equation} \label{eq:basis}
    \begin{split}
        \ket{++} & = \frac{1}{2} \begin{pmatrix} +1 & +1 & +1 & +1 \end{pmatrix}, \\
        \ket{--} & = \frac{1}{2} \begin{pmatrix} +1 & -1 & +1 & -1 \end{pmatrix}, \\
        \ket{+-} & = \frac{1}{2} \begin{pmatrix} +1 & +1 & -1 & -1 \end{pmatrix}, \\
        \ket{-+} & = \frac{1}{2} \begin{pmatrix} +1 & -1 & -1 & +1 \end{pmatrix}.
    \end{split}
\end{equation}

\noindent
Importantly, the basis states $\ket{++}$ and $\ket{--}$ have eigenvalue $+1$ under $\permutation_d$ while the basis states $\ket{+-}$ and $\ket{-+}$ have eigenvalue $-1$ under $\permutation_d$. Hence, the vertex amplitudes, $\ket{\vertexamplitude} = \begin{pmatrix} \vertexamplitude^a & \vertexamplitude^b & \vertexamplitude^c & \vertexamplitude^d \end{pmatrix}$, are divided into sectors that are (anti-)symmetric under such permutations, $\permutation_d \ket{\vertexamplitude_{\pm}} = \pm \ket{\vertexamplitude_{\pm}}$. Substitution of the identity, $\mathbf{1} = \permutation_d \permutation_d$, into the inner product, $\braket{ \vertexamplitude_{\pm} | \compatibility | \vertexamplitude_{\pm} } = \braket{ \pm \vertexamplitude_{\pm} | \permutation_d \compatibility \permutation_d | \pm \vertexamplitude_{\pm} }$, shows that this inner product is equal to its own opposite and hence, it must vanish. Thus, the (anti-)symmetric modes automatically satisfy (anti-)symmetric constraints.

Indeed, the compatibility matrix  constructed from Eq.~(\ref{eq:isometrycondition}) is off-block diagonal (see App.~\ref{appendix:compatibility}) in the basis from Eq.~(\ref{eq:basis}):

\begin{equation} \label{eq:compatibility}
    \compatibility^{\text{sym}} = \begin{pmatrix} 0 & 0 & 0 & 0 \\ 0 & 0 & \globalcoefficient_{24}^- & \globalcoefficient_{13}^- \\ 0 & \globalcoefficient_{24}^- & 0 & 0 \\ 0 & \globalcoefficient_{13}^- & 0 & 0 \end{pmatrix},
\end{equation}
\begin{equation} \label{eq:globalcoefficients}
    \globalcoefficient_i \equiv \frac{ \globaledgedirection_{i+1} \cdot \globaledgedirection_{i+2} \times \globaledgedirection_{i+3} }{ \globaledge_i } \equiv \frac{ \globaledgevector_{i+1} \cdot \globaledgevector_{i+2} \times \globaledgevector_{i+3} }{ \edgeproduct }.
\end{equation}

\noindent
where the $\globalcoefficient_i$ and $\globalcoefficient_j$ are global folding coefficients, $\globalcoefficient_{ij}^- \equiv \globalcoefficient_j - \globalcoefficient_i$ is their difference, and $\edgeproduct \equiv \globaledge_1 \globaledge_2 \globaledge_3 \globaledge_4$ is the product of the four unique edge lengths. In this form, the symmetric and antisymmetric modes in the nullspace are apparent:

\begin{equation} \label{eq:symmetricmode}
    \ket{\vertexamplitude_+} = \ket{++},
\end{equation}
\begin{equation} \label{eq:antisymmetricmode}
    \ket{\vertexamplitude_-} = \globalcoefficient_{13}^- \ket{+-} - \globalcoefficient_{24}^- \ket{-+},
\end{equation}

\noindent
and the corresponding face amplitudes are determined explicitly in App.~\ref{appendix:compatibility}. In fact, the symmetric mode in Eq.~(\ref{eq:symmetricmode}) is the rigid folding motion that generates the degenerate ground states shown in Fig.~\ref{fig:system}B. The modes in Eqs.~(\ref{eq:symmetricmode},\ref{eq:antisymmetricmode}) along with the uniform face-bending mode, $\ket{\vertexamplitude} = \mathbf{0}$, span the three-dimensional space of linear isometries for four-parallelogram origami.

\subsection*{Strain and curvature}

Linear isometries stretch and rotate the lattice vectors which are respectively captured via the strain of the cell and the curvature between cells. Here, we derive these quantities for four-parallelogram origami as linear combinations of the vertex amplitudes and show that they have opposite eigenvalues under the permutation operator $\permutation_d$. We then show that this decouples the in-plane and out-of-plane modes and ensures that the corresponding Poisson's ratios are equal and opposite.

As measured at corner $(a,A)$, the lattice vector deforms according to Eq.~(\ref{eq:vertexdisplacement}) as $\latticevectorchange_\mu^{(a,A)} \equiv \vertexdisplacement^{(a,A)}(\mu=1) - \vertexdisplacement^{(a,A)}(\mu=0)$ so that its change in length must be quadratic in the global folding coefficients. Conveniently, the torsion displaces the lattice vector along $\localedgevector_i^A \times \localedgevector_i^{A'}$ and the additional folding between the corners of a single vertex displaces the lattice vector along $\localedgedirection_i^a \times \latticevector_{\mu}$ which both vanish when projected onto the original lattice vector, $\latticevector_{\mu} = \localedgevector_i^A + \localedgevector_i^{A'}$ (see App.~\ref{appendix:deformations}). Thus, the lattice vector stretches locally depend only on the vertex amplitudes:

\begin{align}
    \latticevector_1 \cdot \latticevectorchange_1^{(a,A)}  & = \edgeproduct \globalcoefficient_2 \globalcoefficient_4 \braket{a | \permutation_h | \vertexamplitude}, \label{eq:stretch1} \\
    \latticevector_2 \cdot \latticevectorchange_2^{(a,A)} & = - \edgeproduct \globalcoefficient_1 \globalcoefficient_3 \braket{a | \permutation_v | \vertexamplitude}, \label{eq:stretch2}
\end{align}

\noindent
where the bra $\bra{a}$ is the original basis vector, $\braket{a | \vertexamplitude} = \vertexamplitude^a$, for the vertex the stretch is measured at.

\begin{figure}
    \centering
    \includegraphics{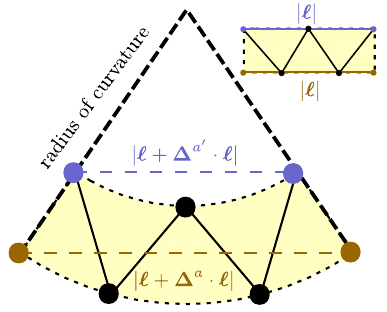}
    \caption{An illustration of the slab-like behavior of corrugated origami sheets. Bending motions extend/compress the lattice vectors, $\latticevectorchange \cdot \latticevector$, by different amounts depending on which vertex, $a$, they are measured from so that the radius of curvature induced by a linear isometry is proportional gradients in the lattice vectors stretches.}
    \label{fig:slab}
\end{figure}

While origami are intrinsically sheets, their corrugation implies that they deform as an elastic slab: upwards bending extends the top, contracts the bottom, and preserves lengths in the mid-plane as illustrated in Fig~\ref{fig:slab}. This intuition explains the vertex-dependence on the lattice vector stretches in Eqs.~(\ref{eq:stretch1},~\ref{eq:stretch2}) as an analog of the height dependence on the strain of the slab. Thus, the strain of the origami sheet can be specified via that of its mid-plane by averaging the lattice vector stretches over its $N_v = 4$ vertices:

\begin{equation} \label{eq:latticestrain}
    \latticestrain_{\mu \mu} \equiv \frac{  \latticevector_{\mu} \cdot \Big( \sum_{a'} \latticevectorchange_{\mu}^{a'} \Big)}{N_v} \equiv \braket{\latticestrain_{\mu \mu} | \vertexamplitude}.
\end{equation}

\noindent
Note that these strains have units of area because the surface coordinates are the dimensionless cell indices and that the subscripts denote the lattice directions without implied summation.

This same connection to elastic slabs suggests that the curvature of the sheet can also be characterized by the vertex-dependence of the lattice vector stretches. The lattice vectors are rotated between cells by the corner-independent lattice angular velocity, $\latticerotation_{\mu} \equiv \angularvelocity^{(a,A)}(\mu = 1) - \angularvelocity^{(a,A)}(\mu = 0)$, which lies in the plane of the sheet with normal vector $\normalvector = \latticevector_1 \times \latticevector_2$ to satisfy the compatibility condition in~Eq.~(\ref{eq:closure}). Compatibility of the vertex displacements in Eq.~(\ref{eq:vertexdisplacement}) shows that changes to the lattice vectors vary between two corners, $\latticevectorchange_{\mu}^{(a,A)} - \latticevectorchange_{\mu}^{(a',A')}$, due to the lattice rotation of the vector between the two vertices, $\latticerotation_{\mu} \times \globaledgevector_{(a,a')}$, and due to the rotation of the lattice vector by the angular velocity gradient between the two corners, $( \angularvelocity^{(a',A')} - \angularvelocity^{(a,A)} ) \times \latticevector_{\mu}$ (see App.~\ref{appendix:deformations}). The latter vanishes under projection onto the initial lattice vector while the former rearranges to define the curvature, $\latticecurvature_{\mu \mu} = \latticerotation_{\mu} \times \latticevector_{\mu} \cdot \normalvector$, via differences in the lattice vector stretches:

\begin{equation} \label{eq:latticecurvature}
    \latticecurvature_{\mu \mu} = -\frac{ \latticevector_{\mu} \cdot \big( \latticevectorchange_{\mu}^{a'} - \latticevectorchange_{\mu}^a \big) }{\globaledgevector_{(a,a')} \cdot \normalvector }\equiv \braket{\latticecurvature_{\mu \mu} | \vertexamplitude}. 
\end{equation}

\noindent
Note that these curvatures are vertex independent and, have units of inverse length because the normal vector has units of area.

In this form, both the lattice strains and the lattice curvatures depend explicitly on the locally-computed lattice vector stretches. However, these operators have opposite eigenvalues under permutations of the non-adjacent vertices,

\begin{align}
\permutation_d \ket{\latticestrain_{\mu \mu}} & = + \ket{\latticestrain_{\mu \mu}}, \\ \permutation_d \ket{\latticecurvature_{\mu \mu}} & = - \ket{\latticecurvature_{\mu \mu}},
\end{align}

\noindent 
hence, they couple to opposite sectors of linear isometries. In fact, since each sector contains exactly one mode, the in-plane strains are exclusively generated by the symmetric mode, $\ket{\vertexamplitude_+}$, and the out-of-plane curvatures are exclusively generated by the antisymmetric mode, $\ket{\vertexamplitude_-}$, which are orthogonal to one another. Thus, the fundamental symmetries explain why these structures have one mode that generates only in-plane strain and one mode that generates only out-of-plane bending, as observed previously in the planar and bend modes of previous work~\cite{schenk2013geometry,wei2013geometric,pratapa2019geometric}.

Lastly, the relative strain and curvature in transverse lattice directions define geometric Poisson's ratios: 

\begin{align}
\poisson_{\text{in}} & \equiv - \frac{\lvert \latticevector_2 \rvert^2}{\lvert \latticevector_1 \rvert^2} \frac{\latticestrain_{11}}{ \latticestrain_{22}}, \\
\poisson_{\text{out}} & \equiv - \frac{\lvert \latticevector_2 \rvert^2}{\lvert \latticevector_1 \rvert^2} \frac{\latticecurvature_{11}}{ \latticecurvature_{22}}.
\end{align} 

\noindent
Such ratios depend entirely on the triple products entering the lattice vector stretches with the exception that the antisymmetry of the lattice curvature operators leads to opposite signs in the numerator and denominator while the symmetry of the lattice strain operators leads to the same signs. Thus, the Poisson's ratios are always equal and opposite:

\begin{equation} \label{eq:poisson}
    \poisson_{\text{in}} = - \poisson_{\text{out}} = \frac{\lvert \latticevector_2 \rvert^2}{\lvert \latticevector_1 \rvert^2} \frac{\globalcoefficient_2 \globalcoefficient_4}{\globalcoefficient_1 \globalcoefficient_3}.
\end{equation}

\noindent
Note that this Poisson's ratio is a purely geometric quantity and it can increase and decrease without limit because the lattice is not isotropic~\cite{ting2005poisson}. Furthermore, recall that generic four-parallelogram origami sheets are not orthotropic so that their Poisson's ratios are not necessarily equal and opposite as measured from orthogonal directions.

The off-diagonal components of the strain and curvature are computed in App.~\ref{appendix:deformations} and exhibit two important features: (i) the lattice shears, $\latticestrain_{12} = \latticestrain_{21} = 0$, always vanish because the inner product $\latticevector_1 \cdot \latticevector_2$ depends only on the length of the edges and the interior angles of the faces and (ii) the uniform face-bending mode exclusively generates twisting characterized by non-vanishing lattice curvatures, $\latticecurvature_{12} = \latticecurvature_{21}$, as previously observed for the special case of the Miura-ori~\cite{wei2013geometric,schenk2013geometry,evans2015lattice}. Together with the diagonal components, these quantities correspond to discrete analogs of the first and second fundamental forms~\cite{lovelock1989tensors} as discussed in App.~\ref{appendix:forms}. 

\subsection*{Poisson's ratio transitions}

The Poisson’s ratios in Eq.~(\ref{eq:poisson}) change along the one-dimensional manifold of degenerate ground states; in particular, the Poisson's ratios can undergo transitions from positive to negative and vice versa as identified in the Morph family~\cite{pratapa2019geometric}. Here, we show that such transitions occur whenever two adjacent faces are co-planar and determine the relationship between the intrinsic crease geometry and such ground states to reveal two subsets of four-parallelogram origami which are strictly in-plane auxetic and strictly out-of-plane auxetic.

\begin{figure}
    \centering
    \includegraphics{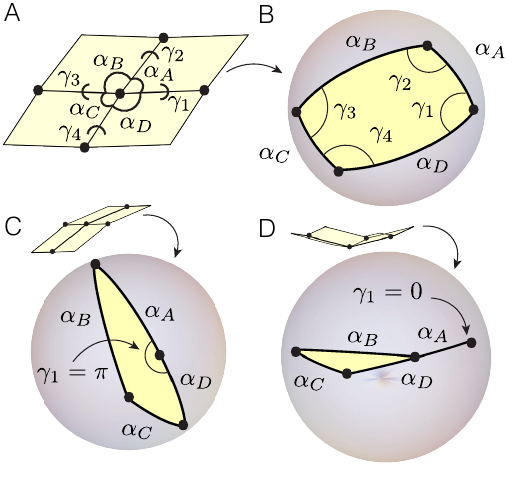}
    \caption{(A) A generic four-coordinated vertex whose geometry is characterized by the sector angles, $\sectorangle$, between subsequent edges and its ground state specified by the dihedral angles, $\dihedralangle$, between adjacent faces. (B) The spherical quadrilateral obtained by projecting the edges in panel A onto the unit sphere. (C) ``Opened'' ($\dihedralangle_1 = \pi$) and (D) ``closed'' ($\dihedralangle_1 = 0$) face configurations whose sector angles respectively satisfy Eq.~(\ref{eq:opened}) and Eq.~(\ref{eq:closed}).}
    \label{fig:projections}
\end{figure}

The Poisson’s ratios change sign when the ratio of triple products in Eq.~(\ref{eq:poisson}) changes sign, which  occurs when three of the edge vectors lie in the same plane. This corresponds to a ground state where two adjacent faces either lie flat to (see inset of Fig.~\ref{fig:projections}C) or lie on top of (see inset of Fig.~\ref{fig:projections}D) one another which are respectively quantified by the dihedral angles between the faces: $\dihedralangle_i = \pi$ or $\dihedralangle_i = 0, 2\pi$. The spatial symmetry of the crease pattern implies that there are only four unique dihedral angles, as shown in Fig.~\ref{fig:projections}A, and that those on parallel edges are complementary to one another, $2\pi - \dihedralangle_i$. Hence, the condition for the Poisson’s ratios to change sign is determined by the existence of ground states for a single vertex with exactly one dihedral angle equal to an integer multiple of $\pi$.

The ground states of an origami vertex are the sets of dihedral angles compatible with the fixed sector angles, $\sectorangle_A$, (see Refs.~\cite{huffman1976curvature,hull2002modelling}) labeled in Fig.~\ref{fig:projections}A. The edges emanating from this vertex in a generic ground state project onto the unit sphere, thereby mapping to the vertices of the spherical quadrilateral shown in Fig.~\ref{fig:projections}B. This spherical quadrilateral has edges that are segments of great circles (geodesics of the sphere) with arclength equal to the corresponding sector angles and interior angles that are equal to the corresponding dihedral angles. This mapping can be used to parameterize the ground states of a generic four-coordinated vertex via spherical trigonometry~\cite{huffman1976curvature,berry2020topological}, (see App.~\ref{appendix:system}).

For the sake of determining the relationship between sector angles and Poisson’s ratio transitions in four-parallelogram origami, it suffices to determine the conditions for the ``opened'' configuration, $\dihedralangle_1 = \pi$, shown in Fig.~\ref{fig:projections}C, and the ``closed'' configuration, $\dihedralangle_1 = 0$, shown in Fig.~\ref{fig:projections}D. In both cases, the spherical quadrilateral flattens to a spherical triangle; hence, this dihedral angle, $\dihedralangle_1$, can take the value of $\pi$ and $0$ only when the respective spherical triangle inequalities are satisfied:

\begin{align}
\spherenorm{ \sectorangle_A + \sectorangle_D } \leq \spherenorm{ \sectorangle_B + \sectorangle_C }, \label{eq:opened} \\
\spherenorm{ \sectorangle_A - \sectorangle_D } \geq \spherenorm{ \sectorangle_B - \sectorangle_C }, \label{eq:closed}
\end{align}

\noindent
where the spherical norm, $\spherenorm{x} \equiv \min\{ x, 2\pi - x \}$, takes the shortest of the two great circles connecting the vertices. Note that when one of the inequalities in Eqs.~(\ref{eq:opened},~\ref{eq:closed}) is not satisfied, the corresponding inequality for $\dihedralangle_3$ \emph{is} satisfied.

Generic choices of sector angles satisfy either one or both of the inequalities in Eqs.~(\ref{eq:opened},~\ref{eq:closed}) for a particular crease. However, the constraints on the dihedral angles always possess two solutions (which are related by the inversion of the sheet $\dihedralangle_i \rightarrow 2\pi - \dihedralangle_i$) and these two cases correspond to different configuration space topologies~\cite{chiang1984classification}. In the former, the two branches connect to satisfy the $2 \pi$ periodicity of the constraints as shown in Fig.~\ref{fig:projections}A. In the latter the two branches are disconnected because each branch is itself $2 \pi$ periodic as shown in Fig.~\ref{fig:projections}B. Note that this feature is independent of flattened configurations~\cite{chen2018branches,mcinerney2020hidden} or curvature of the vertex~\cite{berry2020topological}. Nonetheless, both cases generically undergo Poisson’s ratio transitions as shown in Fig.~\ref{fig:projections}A-B where it is important to note transitions at $\dihedralangle_i = 0, 2\pi$ require self-intersection of the origami sheet and are hence non-physical.

\begin{figure*}
    \centering
    \includegraphics{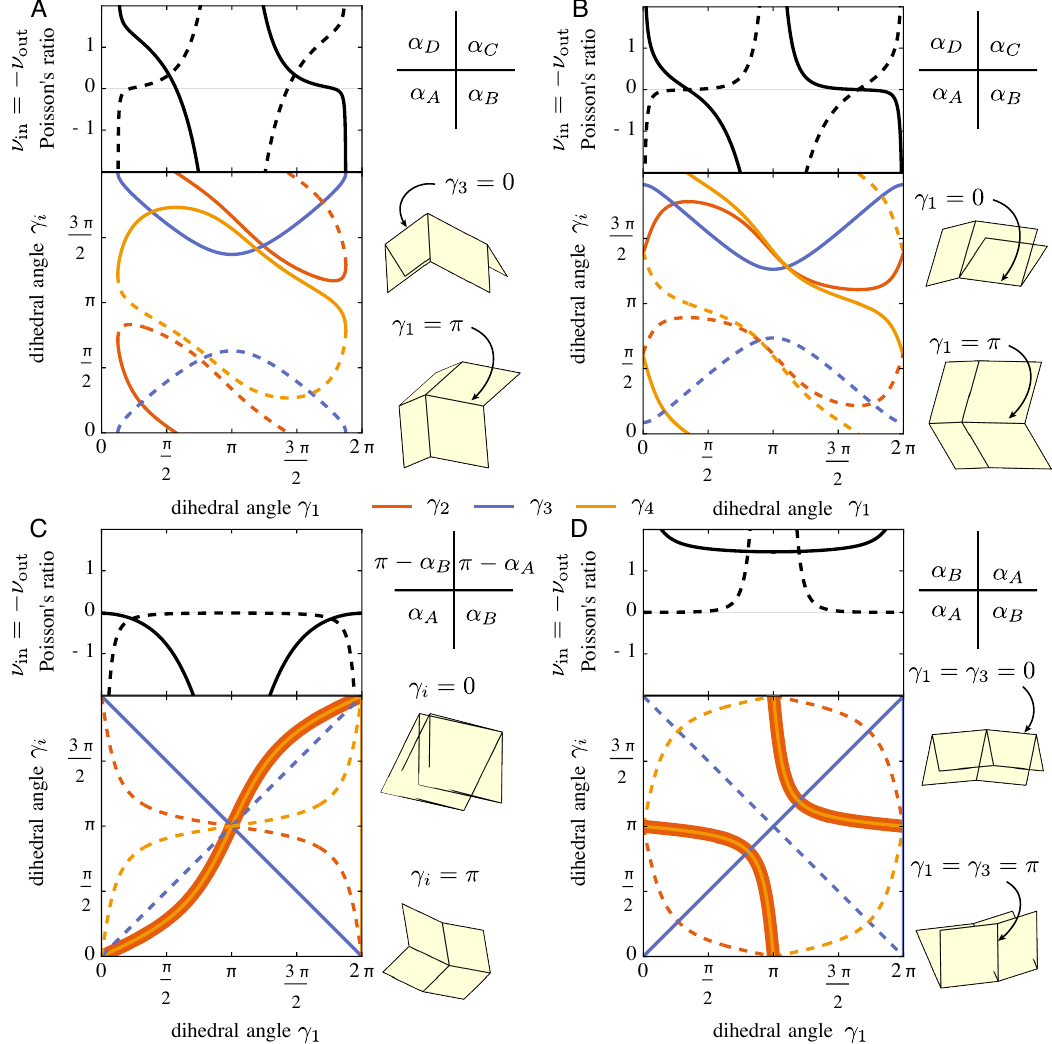}
    \caption{The Poisson's ratios, $\poisson$, and one-dimensional manifolds of degenerate ground states, $\dihedralangle_i = \dihedralangle_i(\dihedralangle_1)$, for four-parallelogram origami sheets with (A) generic sector angles and connected solution branches, (B) generic sector angles and disconnected solution branches, (C) sector angles that ensure a strictly negative in-plane Poisson's ratio, and (D) sector angles that ensure a strictly positive in-plane Poisson's ratio. The Poisson's ratios change sign when exactly two of the faces are co-planar as indicated by a single dihedral angle folding through $0$ or $\pi$.}
    \label{fig:poissons}
\end{figure*}

Special choices of sector angles that satisfy \emph{equality} for both spherical inequalities have ground states for which the dihedral angles fold through $0$ or $\pi$ simultaneously. First, consider developable crease patterns, $\sum \sectorangle_A = 2\pi$, satisfying the Kawasaki condition for flat foldability~\cite{o2011fold}, $\sectorangle_A + \sectorangle_C = \sectorangle_B + \sectorangle_D$, such as that shown in Fig.~\ref{fig:poissons}C. In this case, every dihedral angle folds through $0$ or $\pi$ simultaneously, thereby preventing the Poisson’s ratio to change sign. Second, consider crease patterns satisfying the generalized flat-foldable condition identified in Ref.~\cite{waitukaitis2015origami}, $\sectorangle_A = \sectorangle_C$, $\sectorangle_B = \sectorangle_D$, such as that shown in Fig.~\ref{fig:poissons}D. In this case, non-adjacent dihedral angles fold through $\dihedralangle_i = \dihedralangle_{i+2} = 0$ or $\pi$ simultaneously while the remaining two dihedral angle angles fold through $\dihedralangle_{i+1} = \dihedralangle_{i+3} = \pi$ or $0$ respectively, thereby preventing the Poisson’s ratio to change sign. These two subsets of in-plane and out-of-plane negative Poisson’s ratio origami metamaterials respectively reduce to the Miura-ori~\cite{schenk2013geometry,wei2013geometric} and eggbox~\cite{nassar2017curvature} crease patterns.

\section*{Concluding remarks}

We have developed a framework for characterizing the low-energy deformations of quadrilateral-based origami sheets that enables investigations of the interplay between discrete symmetries and mechanical response. We have implemented this formalism for periodic crease patterns with unit cells composed of four generic parallelogram faces, thereby revealing that in such sheets the infinitesimal strains are induced by a symmetric mode whereas the infinitesimal curvatures are induced by an antisymmetric mode. Moreover, we have showed that these quantities define geometric Poisson's ratios that are always equal and opposite to one another, contrasting the relationship found in conventional elasticity. Lastly, we have discovered subsets within this family of crease patterns that have strictly negative in-plane or out-of-plane Poisson's ratios which may be utilized as auxetic mechanical metamaterials.

The formalism developed herein extends to the analysis of crease patterns beyond four-parallelogram origami. The compatibility conditions can be immediately applied to determine the linear isometries in parallelogram-based origami with larger unit cells~\cite{klett2013realtime,zhou2016origami} as well as Bloch-periodic modes~\cite{evans2015lattice,mcinerney2020hidden}. Furthermore, the symmetry analysis of these compatibility conditions could be used to constrain the stiffness of parallel-based origami tubes~\cite{overvelde2016three,filipov2015origami}. Lastly, the compatibility conditions can also be used to explore the role of discrete symmetries in crease patterns with more generic quadrilateral faces where the face amplitudes cannot be integrated out of the linear constraints~\cite{tachi2009generalization,waitukaitis2015origami,evans2015rigidly,feng2020designs,dieleman2020jigsaw}.

The experimental realization of these structures requires a number of important considerations. The work in the present manuscript considers only the uniform \emph{geometry} of the origami sheets whereas the elastic properties of physical materials can favor one mode of deformation over the others~\cite{schenk2010origami,filipov2017bar,liu2017nonlinear} and imperfections can significantly alter the mechanical response~\cite{pinson2017self,liu2020big}. Moreover, local actuation typically leads to nonuniform deformations~\cite{stern2017complexity,grey2018local,grey2019strain} and creases tend to exhibit temporal relaxation dynamics~\cite{thiria2011relaxation} that give rise to plastic memory effects~\cite{jules2020plasticity}. Thus, it remains to explore the ability to control the mechanical response of physical origami via this geometric symmetry.

In conclusion, the symmetry analysis of linear isometries provides a systematic characterization of the large scale deformations in parallelogram-based origami. More generally, similar analysis can be applied to systems where symmetries constrain both the intracellular and intercellular properties which may be insightful for the design of metamaterials in mechanics~\cite{zadpoor2016mechanical,bertoldi2017flexible,yu2018mechanical,barchiesi2019mechanical,surjadi2019mechanical} and beyond~\cite{wegener2013metamaterials,kadic2013metamaterials}.

\begin{acknowledgments}
The authors thank Michael Czajkowski for helpful discussions and Christian D. Santangelo for feedback on the manuscript.
\end{acknowledgments}

\bibliography{references.bib}

\begin{thebibliography}{81}%
\makeatletter
\providecommand \@ifxundefined [1]{%
 \@ifx{#1\undefined}
}%
\providecommand \@ifnum [1]{%
 \ifnum #1\expandafter \@firstoftwo
 \else \expandafter \@secondoftwo
 \fi
}%
\providecommand \@ifx [1]{%
 \ifx #1\expandafter \@firstoftwo
 \else \expandafter \@secondoftwo
 \fi
}%
\providecommand \natexlab [1]{#1}%
\providecommand \enquote  [1]{``#1''}%
\providecommand \bibnamefont  [1]{#1}%
\providecommand \bibfnamefont [1]{#1}%
\providecommand \citenamefont [1]{#1}%
\providecommand \href@noop [0]{\@secondoftwo}%
\providecommand \href [0]{\begingroup \@sanitize@url \@href}%
\providecommand \@href[1]{\@@startlink{#1}\@@href}%
\providecommand \@@href[1]{\endgroup#1\@@endlink}%
\providecommand \@sanitize@url [0]{\catcode `\\12\catcode `\$12\catcode
  `\&12\catcode `\#12\catcode `\^12\catcode `\_12\catcode `\%12\relax}%
\providecommand \@@startlink[1]{}%
\providecommand \@@endlink[0]{}%
\providecommand \url  [0]{\begingroup\@sanitize@url \@url }%
\providecommand \@url [1]{\endgroup\@href {#1}{\urlprefix }}%
\providecommand \urlprefix  [0]{URL }%
\providecommand \Eprint [0]{\href }%
\providecommand \doibase [0]{https://doi.org/}%
\providecommand \selectlanguage [0]{\@gobble}%
\providecommand \bibinfo  [0]{\@secondoftwo}%
\providecommand \bibfield  [0]{\@secondoftwo}%
\providecommand \translation [1]{[#1]}%
\providecommand \BibitemOpen [0]{}%
\providecommand \bibitemStop [0]{}%
\providecommand \bibitemNoStop [0]{.\EOS\space}%
\providecommand \EOS [0]{\spacefactor3000\relax}%
\providecommand \BibitemShut  [1]{\csname bibitem#1\endcsname}%
\let\auto@bib@innerbib\@empty
\bibitem [{\citenamefont {Weyl}(1950)}]{weyl1950theory}%
  \BibitemOpen
  \bibfield  {author} {\bibinfo {author} {\bibfnamefont {H.}~\bibnamefont
  {Weyl}},\ }\href@noop {} {\emph {\bibinfo {title} {The theory of groups and
  quantum mechanics}}}\ (\bibinfo  {publisher} {Courier Corporation},\ \bibinfo
  {year} {1950})\BibitemShut {NoStop}%
\bibitem [{\citenamefont {Dresselhaus}\ \emph {et~al.}(2007)\citenamefont
  {Dresselhaus}, \citenamefont {Dresselhaus},\ and\ \citenamefont
  {Jorio}}]{dresselhaus2007group}%
  \BibitemOpen
  \bibfield  {author} {\bibinfo {author} {\bibfnamefont {M.~S.}\ \bibnamefont
  {Dresselhaus}}, \bibinfo {author} {\bibfnamefont {G.}~\bibnamefont
  {Dresselhaus}},\ and\ \bibinfo {author} {\bibfnamefont {A.}~\bibnamefont
  {Jorio}},\ }\href@noop {} {\emph {\bibinfo {title} {Group theory: application
  to the physics of condensed matter}}}\ (\bibinfo  {publisher} {Springer
  Science \& Business Media},\ \bibinfo {year} {2007})\BibitemShut {NoStop}%
\bibitem [{\citenamefont {Altland}\ and\ \citenamefont
  {Zirnbauer}(1997)}]{altland1997nonstandard}%
  \BibitemOpen
  \bibfield  {author} {\bibinfo {author} {\bibfnamefont {A.}~\bibnamefont
  {Altland}}\ and\ \bibinfo {author} {\bibfnamefont {M.~R.}\ \bibnamefont
  {Zirnbauer}},\ }\bibfield  {title} {\bibinfo {title} {Nonstandard symmetry
  classes in mesoscopic normal-superconducting hybrid structures},\ }\href@noop
  {} {\bibfield  {journal} {\bibinfo  {journal} {Physical Review B}\ }\textbf
  {\bibinfo {volume} {55}},\ \bibinfo {pages} {1142} (\bibinfo {year}
  {1997})}\BibitemShut {NoStop}%
\bibitem [{\citenamefont {Schnyder}\ \emph {et~al.}(2008)\citenamefont
  {Schnyder}, \citenamefont {Ryu}, \citenamefont {Furusaki},\ and\
  \citenamefont {Ludwig}}]{schnyder2008classification}%
  \BibitemOpen
  \bibfield  {author} {\bibinfo {author} {\bibfnamefont {A.~P.}\ \bibnamefont
  {Schnyder}}, \bibinfo {author} {\bibfnamefont {S.}~\bibnamefont {Ryu}},
  \bibinfo {author} {\bibfnamefont {A.}~\bibnamefont {Furusaki}},\ and\
  \bibinfo {author} {\bibfnamefont {A.~W.}\ \bibnamefont {Ludwig}},\ }\bibfield
   {title} {\bibinfo {title} {Classification of topological insulators and
  superconductors in three spatial dimensions},\ }\href@noop {} {\bibfield
  {journal} {\bibinfo  {journal} {Physical Review B}\ }\textbf {\bibinfo
  {volume} {78}},\ \bibinfo {pages} {195125} (\bibinfo {year}
  {2008})}\BibitemShut {NoStop}%
\bibitem [{\citenamefont {Kitaev}(2009)}]{kitaev2009periodic}%
  \BibitemOpen
  \bibfield  {author} {\bibinfo {author} {\bibfnamefont {A.}~\bibnamefont
  {Kitaev}},\ }\bibfield  {title} {\bibinfo {title} {Periodic table for
  topological insulators and superconductors},\ }in\ \href@noop {} {\emph
  {\bibinfo {booktitle} {AIP conference proceedings}}},\ Vol.\ \bibinfo
  {volume} {1134}\ (\bibinfo {organization} {American Institute of Physics},\
  \bibinfo {year} {2009})\ pp.\ \bibinfo {pages} {22--30}\BibitemShut {NoStop}%
\bibitem [{\citenamefont {Ryu}\ \emph {et~al.}(2010)\citenamefont {Ryu},
  \citenamefont {Schnyder}, \citenamefont {Furusaki},\ and\ \citenamefont
  {Ludwig}}]{ryu2010topological}%
  \BibitemOpen
  \bibfield  {author} {\bibinfo {author} {\bibfnamefont {S.}~\bibnamefont
  {Ryu}}, \bibinfo {author} {\bibfnamefont {A.~P.}\ \bibnamefont {Schnyder}},
  \bibinfo {author} {\bibfnamefont {A.}~\bibnamefont {Furusaki}},\ and\
  \bibinfo {author} {\bibfnamefont {A.~W.}\ \bibnamefont {Ludwig}},\ }\bibfield
   {title} {\bibinfo {title} {Topological insulators and superconductors:
  tenfold way and dimensional hierarchy},\ }\href@noop {} {\bibfield  {journal}
  {\bibinfo  {journal} {New Journal of Physics}\ }\textbf {\bibinfo {volume}
  {12}},\ \bibinfo {pages} {065010} (\bibinfo {year} {2010})}\BibitemShut
  {NoStop}%
\bibitem [{\citenamefont {Roychowdhury}\ and\ \citenamefont
  {Lawler}(2018)}]{roychowdhury2018classification}%
  \BibitemOpen
  \bibfield  {author} {\bibinfo {author} {\bibfnamefont {K.}~\bibnamefont
  {Roychowdhury}}\ and\ \bibinfo {author} {\bibfnamefont {M.~J.}\ \bibnamefont
  {Lawler}},\ }\bibfield  {title} {\bibinfo {title} {Classification of magnetic
  frustration and metamaterials from topology},\ }\href@noop {} {\bibfield
  {journal} {\bibinfo  {journal} {Physical Review B}\ }\textbf {\bibinfo
  {volume} {98}},\ \bibinfo {pages} {094432} (\bibinfo {year}
  {2018})}\BibitemShut {NoStop}%
\bibitem [{\citenamefont {S{\"u}sstrunk}\ and\ \citenamefont
  {Huber}(2016)}]{susstrunk2016classification}%
  \BibitemOpen
  \bibfield  {author} {\bibinfo {author} {\bibfnamefont {R.}~\bibnamefont
  {S{\"u}sstrunk}}\ and\ \bibinfo {author} {\bibfnamefont {S.~D.}\ \bibnamefont
  {Huber}},\ }\bibfield  {title} {\bibinfo {title} {Classification of
  topological phonons in linear mechanical metamaterials},\ }\href@noop {}
  {\bibfield  {journal} {\bibinfo  {journal} {Proceedings of the National
  Academy of Sciences}\ }\textbf {\bibinfo {volume} {113}},\ \bibinfo {pages}
  {E4767} (\bibinfo {year} {2016})}\BibitemShut {NoStop}%
\bibitem [{\citenamefont {Guest}\ and\ \citenamefont
  {Fowler}(2014)}]{guest2014}%
  \BibitemOpen
  \bibfield  {author} {\bibinfo {author} {\bibfnamefont {S.}~\bibnamefont
  {Guest}}\ and\ \bibinfo {author} {\bibfnamefont {P.}~\bibnamefont {Fowler}},\
  }\bibfield  {title} {\bibinfo {title} {Symmetry-extended counting rules for
  periodic frameworks},\ }\href@noop {} {\bibfield  {journal} {\bibinfo
  {journal} {Philosophical Transactions of the Royal Society A: Mathematical,
  Physical and Engineering Sciences}\ }\textbf {\bibinfo {volume} {372}},\
  \bibinfo {pages} {20120029} (\bibinfo {year} {2014})}\BibitemShut {NoStop}%
\bibitem [{\citenamefont {Mitschke}\ \emph {et~al.}(2013)\citenamefont
  {Mitschke}, \citenamefont {Schr{\"o}der-Turk}, \citenamefont {Mecke},
  \citenamefont {Fowler},\ and\ \citenamefont {Guest}}]{mitschke2013symmetry}%
  \BibitemOpen
  \bibfield  {author} {\bibinfo {author} {\bibfnamefont {H.}~\bibnamefont
  {Mitschke}}, \bibinfo {author} {\bibfnamefont {G.}~\bibnamefont
  {Schr{\"o}der-Turk}}, \bibinfo {author} {\bibfnamefont {K.}~\bibnamefont
  {Mecke}}, \bibinfo {author} {\bibfnamefont {P.}~\bibnamefont {Fowler}},\ and\
  \bibinfo {author} {\bibfnamefont {S.}~\bibnamefont {Guest}},\ }\bibfield
  {title} {\bibinfo {title} {Symmetry detection of auxetic behaviour in 2d
  frameworks},\ }\href@noop {} {\bibfield  {journal} {\bibinfo  {journal} {EPL
  (Europhysics Letters)}\ }\textbf {\bibinfo {volume} {102}},\ \bibinfo {pages}
  {66005} (\bibinfo {year} {2013})}\BibitemShut {NoStop}%
\bibitem [{\citenamefont {Fowler}\ \emph {et~al.}(2014)\citenamefont {Fowler},
  \citenamefont {Guest},\ and\ \citenamefont {Tarnai}}]{fowler2014symmetry}%
  \BibitemOpen
  \bibfield  {author} {\bibinfo {author} {\bibfnamefont {P.~W.}\ \bibnamefont
  {Fowler}}, \bibinfo {author} {\bibfnamefont {S.~D.}\ \bibnamefont {Guest}},\
  and\ \bibinfo {author} {\bibfnamefont {T.}~\bibnamefont {Tarnai}},\
  }\bibfield  {title} {\bibinfo {title} {Symmetry perspectives on some auxetic
  body-bar frameworks},\ }\href@noop {} {\bibfield  {journal} {\bibinfo
  {journal} {Symmetry}\ }\textbf {\bibinfo {volume} {6}},\ \bibinfo {pages}
  {368} (\bibinfo {year} {2014})}\BibitemShut {NoStop}%
\bibitem [{\citenamefont {Babaee}\ \emph {et~al.}(2013)\citenamefont {Babaee},
  \citenamefont {Shim}, \citenamefont {Weaver}, \citenamefont {Chen},
  \citenamefont {Patel},\ and\ \citenamefont {Bertoldi}}]{babaee20133d}%
  \BibitemOpen
  \bibfield  {author} {\bibinfo {author} {\bibfnamefont {S.}~\bibnamefont
  {Babaee}}, \bibinfo {author} {\bibfnamefont {J.}~\bibnamefont {Shim}},
  \bibinfo {author} {\bibfnamefont {J.~C.}\ \bibnamefont {Weaver}}, \bibinfo
  {author} {\bibfnamefont {E.~R.}\ \bibnamefont {Chen}}, \bibinfo {author}
  {\bibfnamefont {N.}~\bibnamefont {Patel}},\ and\ \bibinfo {author}
  {\bibfnamefont {K.}~\bibnamefont {Bertoldi}},\ }\bibfield  {title} {\bibinfo
  {title} {3d soft metamaterials with negative poisson's ratio},\ }\href@noop
  {} {\bibfield  {journal} {\bibinfo  {journal} {Advanced Materials}\ }\textbf
  {\bibinfo {volume} {25}},\ \bibinfo {pages} {5044} (\bibinfo {year}
  {2013})}\BibitemShut {NoStop}%
\bibitem [{\citenamefont {Hou}\ and\ \citenamefont
  {Silberschmidt}(2015)}]{hou2015metamaterials}%
  \BibitemOpen
  \bibfield  {author} {\bibinfo {author} {\bibfnamefont {X.}~\bibnamefont
  {Hou}}\ and\ \bibinfo {author} {\bibfnamefont {V.~V.}\ \bibnamefont
  {Silberschmidt}},\ }\bibfield  {title} {\bibinfo {title} {Metamaterials with
  negative poisson’s ratio: A review of mechanical properties and deformation
  mechanisms},\ }\href@noop {} {\bibfield  {journal} {\bibinfo  {journal}
  {Mechanics of advanced materials}\ ,\ \bibinfo {pages} {155}} (\bibinfo
  {year} {2015})}\BibitemShut {NoStop}%
\bibitem [{\citenamefont {Kolken}\ and\ \citenamefont
  {Zadpoor}(2017)}]{kolken2017auxetic}%
  \BibitemOpen
  \bibfield  {author} {\bibinfo {author} {\bibfnamefont {H.~M.}\ \bibnamefont
  {Kolken}}\ and\ \bibinfo {author} {\bibfnamefont {A.}~\bibnamefont
  {Zadpoor}},\ }\bibfield  {title} {\bibinfo {title} {Auxetic mechanical
  metamaterials},\ }\href@noop {} {\bibfield  {journal} {\bibinfo  {journal}
  {RSC advances}\ }\textbf {\bibinfo {volume} {7}},\ \bibinfo {pages} {5111}
  (\bibinfo {year} {2017})}\BibitemShut {NoStop}%
\bibitem [{\citenamefont {Kelkar}\ \emph {et~al.}(2020)\citenamefont {Kelkar},
  \citenamefont {Kim}, \citenamefont {Cho}, \citenamefont {Kwak}, \citenamefont
  {Kang},\ and\ \citenamefont {Song}}]{kelkar2020cellular}%
  \BibitemOpen
  \bibfield  {author} {\bibinfo {author} {\bibfnamefont {P.~U.}\ \bibnamefont
  {Kelkar}}, \bibinfo {author} {\bibfnamefont {H.~S.}\ \bibnamefont {Kim}},
  \bibinfo {author} {\bibfnamefont {K.-H.}\ \bibnamefont {Cho}}, \bibinfo
  {author} {\bibfnamefont {J.~Y.}\ \bibnamefont {Kwak}}, \bibinfo {author}
  {\bibfnamefont {C.-Y.}\ \bibnamefont {Kang}},\ and\ \bibinfo {author}
  {\bibfnamefont {H.-C.}\ \bibnamefont {Song}},\ }\bibfield  {title} {\bibinfo
  {title} {Cellular auxetic structures for mechanical metamaterials: A
  review},\ }\href@noop {} {\bibfield  {journal} {\bibinfo  {journal}
  {Sensors}\ }\textbf {\bibinfo {volume} {20}},\ \bibinfo {pages} {3132}
  (\bibinfo {year} {2020})}\BibitemShut {NoStop}%
\bibitem [{\citenamefont {Landau}\ and\ \citenamefont
  {Lifshitz}(1959)}]{landau1959course}%
  \BibitemOpen
  \bibfield  {author} {\bibinfo {author} {\bibfnamefont {L.~D.}\ \bibnamefont
  {Landau}}\ and\ \bibinfo {author} {\bibfnamefont {E.~M.}\ \bibnamefont
  {Lifshitz}},\ }\href@noop {} {\emph {\bibinfo {title} {Course of Theoretical
  Physics Vol 7: Theory and Elasticity}}}\ (\bibinfo  {publisher} {Pergamon
  press},\ \bibinfo {year} {1959})\BibitemShut {NoStop}%
\bibitem [{\citenamefont {Wei}\ \emph {et~al.}(2013)\citenamefont {Wei},
  \citenamefont {Guo}, \citenamefont {Dudte}, \citenamefont {Liang},\ and\
  \citenamefont {Mahadevan}}]{wei2013geometric}%
  \BibitemOpen
  \bibfield  {author} {\bibinfo {author} {\bibfnamefont {Z.~Y.}\ \bibnamefont
  {Wei}}, \bibinfo {author} {\bibfnamefont {Z.~V.}\ \bibnamefont {Guo}},
  \bibinfo {author} {\bibfnamefont {L.}~\bibnamefont {Dudte}}, \bibinfo
  {author} {\bibfnamefont {H.~Y.}\ \bibnamefont {Liang}},\ and\ \bibinfo
  {author} {\bibfnamefont {L.}~\bibnamefont {Mahadevan}},\ }\bibfield  {title}
  {\bibinfo {title} {Geometric mechanics of periodic pleated origami},\
  }\href@noop {} {\bibfield  {journal} {\bibinfo  {journal} {Physical review
  letters}\ }\textbf {\bibinfo {volume} {110}},\ \bibinfo {pages} {215501}
  (\bibinfo {year} {2013})}\BibitemShut {NoStop}%
\bibitem [{\citenamefont {Schenk}\ and\ \citenamefont
  {Guest}(2013)}]{schenk2013geometry}%
  \BibitemOpen
  \bibfield  {author} {\bibinfo {author} {\bibfnamefont {M.}~\bibnamefont
  {Schenk}}\ and\ \bibinfo {author} {\bibfnamefont {S.~D.}\ \bibnamefont
  {Guest}},\ }\bibfield  {title} {\bibinfo {title} {Geometry of miura-folded
  metamaterials},\ }\href@noop {} {\bibfield  {journal} {\bibinfo  {journal}
  {Proceedings of the National Academy of Sciences}\ }\textbf {\bibinfo
  {volume} {110}},\ \bibinfo {pages} {3276} (\bibinfo {year}
  {2013})}\BibitemShut {NoStop}%
\bibitem [{\citenamefont {Nassar}\ \emph {et~al.}(2017)\citenamefont {Nassar},
  \citenamefont {Leb{\'e}e},\ and\ \citenamefont
  {Monasse}}]{nassar2017curvature}%
  \BibitemOpen
  \bibfield  {author} {\bibinfo {author} {\bibfnamefont {H.}~\bibnamefont
  {Nassar}}, \bibinfo {author} {\bibfnamefont {A.}~\bibnamefont {Leb{\'e}e}},\
  and\ \bibinfo {author} {\bibfnamefont {L.}~\bibnamefont {Monasse}},\
  }\bibfield  {title} {\bibinfo {title} {Curvature, metric and parametrization
  of origami tessellations: theory and application to the eggbox pattern},\
  }\href@noop {} {\bibfield  {journal} {\bibinfo  {journal} {Proceedings of the
  Royal Society A: Mathematical, Physical and Engineering Sciences}\ }\textbf
  {\bibinfo {volume} {473}},\ \bibinfo {pages} {20160705} (\bibinfo {year}
  {2017})}\BibitemShut {NoStop}%
\bibitem [{\citenamefont {Pratapa}\ \emph {et~al.}(2019)\citenamefont
  {Pratapa}, \citenamefont {Liu},\ and\ \citenamefont
  {Paulino}}]{pratapa2019geometric}%
  \BibitemOpen
  \bibfield  {author} {\bibinfo {author} {\bibfnamefont {P.~P.}\ \bibnamefont
  {Pratapa}}, \bibinfo {author} {\bibfnamefont {K.}~\bibnamefont {Liu}},\ and\
  \bibinfo {author} {\bibfnamefont {G.~H.}\ \bibnamefont {Paulino}},\
  }\bibfield  {title} {\bibinfo {title} {Geometric mechanics of origami
  patterns exhibiting poisson’s ratio switch by breaking mountain and valley
  assignment},\ }\href@noop {} {\bibfield  {journal} {\bibinfo  {journal}
  {Physical review letters}\ }\textbf {\bibinfo {volume} {122}},\ \bibinfo
  {pages} {155501} (\bibinfo {year} {2019})}\BibitemShut {NoStop}%
\bibitem [{\citenamefont {Peraza-Hernandez}\ \emph {et~al.}(2014)\citenamefont
  {Peraza-Hernandez}, \citenamefont {Hartl}, \citenamefont {Malak~Jr},\ and\
  \citenamefont {Lagoudas}}]{peraza2014origami}%
  \BibitemOpen
  \bibfield  {author} {\bibinfo {author} {\bibfnamefont {E.~A.}\ \bibnamefont
  {Peraza-Hernandez}}, \bibinfo {author} {\bibfnamefont {D.~J.}\ \bibnamefont
  {Hartl}}, \bibinfo {author} {\bibfnamefont {R.~J.}\ \bibnamefont
  {Malak~Jr}},\ and\ \bibinfo {author} {\bibfnamefont {D.~C.}\ \bibnamefont
  {Lagoudas}},\ }\bibfield  {title} {\bibinfo {title} {Origami-inspired active
  structures: a synthesis and review},\ }\href@noop {} {\bibfield  {journal}
  {\bibinfo  {journal} {Smart Materials and Structures}\ }\textbf {\bibinfo
  {volume} {23}},\ \bibinfo {pages} {094001} (\bibinfo {year}
  {2014})}\BibitemShut {NoStop}%
\bibitem [{\citenamefont {Rogers}\ \emph {et~al.}(2016)\citenamefont {Rogers},
  \citenamefont {Huang}, \citenamefont {Schmidt},\ and\ \citenamefont
  {Gracias}}]{rogers2016origami}%
  \BibitemOpen
  \bibfield  {author} {\bibinfo {author} {\bibfnamefont {J.}~\bibnamefont
  {Rogers}}, \bibinfo {author} {\bibfnamefont {Y.}~\bibnamefont {Huang}},
  \bibinfo {author} {\bibfnamefont {O.~G.}\ \bibnamefont {Schmidt}},\ and\
  \bibinfo {author} {\bibfnamefont {D.~H.}\ \bibnamefont {Gracias}},\
  }\bibfield  {title} {\bibinfo {title} {Origami mems and nems},\ }\href@noop
  {} {\bibfield  {journal} {\bibinfo  {journal} {Mrs Bulletin}\ }\textbf
  {\bibinfo {volume} {41}},\ \bibinfo {pages} {123} (\bibinfo {year}
  {2016})}\BibitemShut {NoStop}%
\bibitem [{\citenamefont {Santangelo}(2017)}]{santangelo2017extreme}%
  \BibitemOpen
  \bibfield  {author} {\bibinfo {author} {\bibfnamefont {C.~D.}\ \bibnamefont
  {Santangelo}},\ }\bibfield  {title} {\bibinfo {title} {Extreme mechanics:
  self-folding origami},\ }\href@noop {} {\bibfield  {journal} {\bibinfo
  {journal} {Annual Review of Condensed Matter Physics}\ }\textbf {\bibinfo
  {volume} {8}},\ \bibinfo {pages} {165} (\bibinfo {year} {2017})}\BibitemShut
  {NoStop}%
\bibitem [{\citenamefont {Santangelo}(2020)}]{santangelo2020theory}%
  \BibitemOpen
  \bibfield  {author} {\bibinfo {author} {\bibfnamefont {C.~D.}\ \bibnamefont
  {Santangelo}},\ }\bibfield  {title} {\bibinfo {title} {Theory and practice of
  origami in science},\ }\href@noop {} {\bibfield  {journal} {\bibinfo
  {journal} {Soft matter}\ }\textbf {\bibinfo {volume} {16}},\ \bibinfo {pages}
  {94} (\bibinfo {year} {2020})}\BibitemShut {NoStop}%
\bibitem [{\citenamefont {Huffman}(1976)}]{huffman1976curvature}%
  \BibitemOpen
  \bibfield  {author} {\bibinfo {author} {\bibfnamefont {D.~A.}\ \bibnamefont
  {Huffman}},\ }\bibfield  {title} {\bibinfo {title} {Curvature and creases: A
  primer on paper},\ }\href@noop {} {\bibfield  {journal} {\bibinfo  {journal}
  {IEEE Transactions on computers}\ }\textbf {\bibinfo {volume} {25}},\
  \bibinfo {pages} {1010} (\bibinfo {year} {1976})}\BibitemShut {NoStop}%
\bibitem [{\citenamefont {Hull}\ \emph {et~al.}(2002)\citenamefont {Hull} \emph
  {et~al.}}]{hull2002modelling}%
  \BibitemOpen
  \bibfield  {author} {\bibinfo {author} {\bibfnamefont {T.~C.}\ \bibnamefont
  {Hull}} \emph {et~al.},\ }\bibfield  {title} {\bibinfo {title} {Modelling the
  folding of paper into three dimensions using affine transformations},\
  }\href@noop {} {\bibfield  {journal} {\bibinfo  {journal} {Linear Algebra and
  its applications}\ }\textbf {\bibinfo {volume} {348}},\ \bibinfo {pages}
  {273} (\bibinfo {year} {2002})}\BibitemShut {NoStop}%
\bibitem [{\citenamefont {Hawkes}\ \emph {et~al.}(2010)\citenamefont {Hawkes},
  \citenamefont {An}, \citenamefont {Benbernou}, \citenamefont {Tanaka},
  \citenamefont {Kim}, \citenamefont {Demaine}, \citenamefont {Rus},\ and\
  \citenamefont {Wood}}]{hawkes2010programmable}%
  \BibitemOpen
  \bibfield  {author} {\bibinfo {author} {\bibfnamefont {E.}~\bibnamefont
  {Hawkes}}, \bibinfo {author} {\bibfnamefont {B.}~\bibnamefont {An}}, \bibinfo
  {author} {\bibfnamefont {N.~M.}\ \bibnamefont {Benbernou}}, \bibinfo {author}
  {\bibfnamefont {H.}~\bibnamefont {Tanaka}}, \bibinfo {author} {\bibfnamefont
  {S.}~\bibnamefont {Kim}}, \bibinfo {author} {\bibfnamefont {E.~D.}\
  \bibnamefont {Demaine}}, \bibinfo {author} {\bibfnamefont {D.}~\bibnamefont
  {Rus}},\ and\ \bibinfo {author} {\bibfnamefont {R.~J.}\ \bibnamefont
  {Wood}},\ }\bibfield  {title} {\bibinfo {title} {Programmable matter by
  folding},\ }\href@noop {} {\bibfield  {journal} {\bibinfo  {journal}
  {Proceedings of the National Academy of Sciences}\ }\textbf {\bibinfo
  {volume} {107}},\ \bibinfo {pages} {12441} (\bibinfo {year}
  {2010})}\BibitemShut {NoStop}%
\bibitem [{\citenamefont {Tolley}\ \emph {et~al.}(2014)\citenamefont {Tolley},
  \citenamefont {Felton}, \citenamefont {Miyashita}, \citenamefont {Aukes},
  \citenamefont {Rus},\ and\ \citenamefont {Wood}}]{tolley2014self}%
  \BibitemOpen
  \bibfield  {author} {\bibinfo {author} {\bibfnamefont {M.~T.}\ \bibnamefont
  {Tolley}}, \bibinfo {author} {\bibfnamefont {S.~M.}\ \bibnamefont {Felton}},
  \bibinfo {author} {\bibfnamefont {S.}~\bibnamefont {Miyashita}}, \bibinfo
  {author} {\bibfnamefont {D.}~\bibnamefont {Aukes}}, \bibinfo {author}
  {\bibfnamefont {D.}~\bibnamefont {Rus}},\ and\ \bibinfo {author}
  {\bibfnamefont {R.~J.}\ \bibnamefont {Wood}},\ }\bibfield  {title} {\bibinfo
  {title} {Self-folding origami: shape memory composites activated by uniform
  heating},\ }\href@noop {} {\bibfield  {journal} {\bibinfo  {journal} {Smart
  Materials and Structures}\ }\textbf {\bibinfo {volume} {23}},\ \bibinfo
  {pages} {094006} (\bibinfo {year} {2014})}\BibitemShut {NoStop}%
\bibitem [{\citenamefont {Melancon}\ \emph {et~al.}(2021)\citenamefont
  {Melancon}, \citenamefont {Gorissen}, \citenamefont {Garc{\'\i}a-Mora},
  \citenamefont {Hoberman},\ and\ \citenamefont
  {Bertoldi}}]{melancon2021multistable}%
  \BibitemOpen
  \bibfield  {author} {\bibinfo {author} {\bibfnamefont {D.}~\bibnamefont
  {Melancon}}, \bibinfo {author} {\bibfnamefont {B.}~\bibnamefont {Gorissen}},
  \bibinfo {author} {\bibfnamefont {C.~J.}\ \bibnamefont {Garc{\'\i}a-Mora}},
  \bibinfo {author} {\bibfnamefont {C.}~\bibnamefont {Hoberman}},\ and\
  \bibinfo {author} {\bibfnamefont {K.}~\bibnamefont {Bertoldi}},\ }\bibfield
  {title} {\bibinfo {title} {Multistable inflatable origami structures at the
  metre scale},\ }\href@noop {} {\bibfield  {journal} {\bibinfo  {journal}
  {Nature}\ }\textbf {\bibinfo {volume} {592}},\ \bibinfo {pages} {545}
  (\bibinfo {year} {2021})}\BibitemShut {NoStop}%
\bibitem [{\citenamefont {Bassik}\ \emph {et~al.}(2009)\citenamefont {Bassik},
  \citenamefont {Stern},\ and\ \citenamefont
  {Gracias}}]{bassik2009microassembly}%
  \BibitemOpen
  \bibfield  {author} {\bibinfo {author} {\bibfnamefont {N.}~\bibnamefont
  {Bassik}}, \bibinfo {author} {\bibfnamefont {G.~M.}\ \bibnamefont {Stern}},\
  and\ \bibinfo {author} {\bibfnamefont {D.~H.}\ \bibnamefont {Gracias}},\
  }\bibfield  {title} {\bibinfo {title} {Microassembly based on hands free
  origami with bidirectional curvature},\ }\href@noop {} {\bibfield  {journal}
  {\bibinfo  {journal} {Applied physics letters}\ }\textbf {\bibinfo {volume}
  {95}},\ \bibinfo {pages} {091901} (\bibinfo {year} {2009})}\BibitemShut
  {NoStop}%
\bibitem [{\citenamefont {Cho}\ \emph {et~al.}(2011)\citenamefont {Cho},
  \citenamefont {Keung}, \citenamefont {Verellen}, \citenamefont {Lagae},
  \citenamefont {Moshchalkov}, \citenamefont {Van~Dorpe},\ and\ \citenamefont
  {Gracias}}]{cho2011nanoscale}%
  \BibitemOpen
  \bibfield  {author} {\bibinfo {author} {\bibfnamefont {J.-H.}\ \bibnamefont
  {Cho}}, \bibinfo {author} {\bibfnamefont {M.~D.}\ \bibnamefont {Keung}},
  \bibinfo {author} {\bibfnamefont {N.}~\bibnamefont {Verellen}}, \bibinfo
  {author} {\bibfnamefont {L.}~\bibnamefont {Lagae}}, \bibinfo {author}
  {\bibfnamefont {V.}~\bibnamefont {Moshchalkov}}, \bibinfo {author}
  {\bibfnamefont {P.}~\bibnamefont {Van~Dorpe}},\ and\ \bibinfo {author}
  {\bibfnamefont {D.~H.}\ \bibnamefont {Gracias}},\ }\bibfield  {title}
  {\bibinfo {title} {Nanoscale origami for 3d optics},\ }\href@noop {}
  {\bibfield  {journal} {\bibinfo  {journal} {Small}\ }\textbf {\bibinfo
  {volume} {7}},\ \bibinfo {pages} {1943} (\bibinfo {year} {2011})}\BibitemShut
  {NoStop}%
\bibitem [{\citenamefont {Lazarus}\ \emph {et~al.}(2019)\citenamefont
  {Lazarus}, \citenamefont {Smith},\ and\ \citenamefont
  {Dickey}}]{lazarus2019self}%
  \BibitemOpen
  \bibfield  {author} {\bibinfo {author} {\bibfnamefont {N.}~\bibnamefont
  {Lazarus}}, \bibinfo {author} {\bibfnamefont {G.~L.}\ \bibnamefont {Smith}},\
  and\ \bibinfo {author} {\bibfnamefont {M.~D.}\ \bibnamefont {Dickey}},\
  }\bibfield  {title} {\bibinfo {title} {Self-folding metal origami},\
  }\href@noop {} {\bibfield  {journal} {\bibinfo  {journal} {Advanced
  Intelligent Systems}\ }\textbf {\bibinfo {volume} {1}},\ \bibinfo {pages}
  {1900059} (\bibinfo {year} {2019})}\BibitemShut {NoStop}%
\bibitem [{\citenamefont {Liu}\ \emph {et~al.}(2012)\citenamefont {Liu},
  \citenamefont {Boyles}, \citenamefont {Genzer},\ and\ \citenamefont
  {Dickey}}]{liu2012self}%
  \BibitemOpen
  \bibfield  {author} {\bibinfo {author} {\bibfnamefont {Y.}~\bibnamefont
  {Liu}}, \bibinfo {author} {\bibfnamefont {J.~K.}\ \bibnamefont {Boyles}},
  \bibinfo {author} {\bibfnamefont {J.}~\bibnamefont {Genzer}},\ and\ \bibinfo
  {author} {\bibfnamefont {M.~D.}\ \bibnamefont {Dickey}},\ }\bibfield  {title}
  {\bibinfo {title} {Self-folding of polymer sheets using local light
  absorption},\ }\href@noop {} {\bibfield  {journal} {\bibinfo  {journal} {Soft
  matter}\ }\textbf {\bibinfo {volume} {8}},\ \bibinfo {pages} {1764} (\bibinfo
  {year} {2012})}\BibitemShut {NoStop}%
\bibitem [{\citenamefont {Na}\ \emph {et~al.}(2015)\citenamefont {Na},
  \citenamefont {Evans}, \citenamefont {Bae}, \citenamefont {Chiappelli},
  \citenamefont {Santangelo}, \citenamefont {Lang}, \citenamefont {Hull},\ and\
  \citenamefont {Hayward}}]{na2015programming}%
  \BibitemOpen
  \bibfield  {author} {\bibinfo {author} {\bibfnamefont {J.-H.}\ \bibnamefont
  {Na}}, \bibinfo {author} {\bibfnamefont {A.~A.}\ \bibnamefont {Evans}},
  \bibinfo {author} {\bibfnamefont {J.}~\bibnamefont {Bae}}, \bibinfo {author}
  {\bibfnamefont {M.~C.}\ \bibnamefont {Chiappelli}}, \bibinfo {author}
  {\bibfnamefont {C.~D.}\ \bibnamefont {Santangelo}}, \bibinfo {author}
  {\bibfnamefont {R.~J.}\ \bibnamefont {Lang}}, \bibinfo {author}
  {\bibfnamefont {T.~C.}\ \bibnamefont {Hull}},\ and\ \bibinfo {author}
  {\bibfnamefont {R.~C.}\ \bibnamefont {Hayward}},\ }\bibfield  {title}
  {\bibinfo {title} {Programming reversibly self-folding origami with
  micropatterned photo-crosslinkable polymer trilayers},\ }\href@noop {}
  {\bibfield  {journal} {\bibinfo  {journal} {Advanced Materials}\ }\textbf
  {\bibinfo {volume} {27}},\ \bibinfo {pages} {79} (\bibinfo {year}
  {2015})}\BibitemShut {NoStop}%
\bibitem [{\citenamefont {Liu}\ \emph {et~al.}(2016)\citenamefont {Liu},
  \citenamefont {Genzer},\ and\ \citenamefont {Dickey}}]{liu20162d}%
  \BibitemOpen
  \bibfield  {author} {\bibinfo {author} {\bibfnamefont {Y.}~\bibnamefont
  {Liu}}, \bibinfo {author} {\bibfnamefont {J.}~\bibnamefont {Genzer}},\ and\
  \bibinfo {author} {\bibfnamefont {M.~D.}\ \bibnamefont {Dickey}},\ }\bibfield
   {title} {\bibinfo {title} {“2d or not 2d”: Shape-programming polymer
  sheets},\ }\href@noop {} {\bibfield  {journal} {\bibinfo  {journal} {Progress
  in Polymer Science}\ }\textbf {\bibinfo {volume} {52}},\ \bibinfo {pages}
  {79} (\bibinfo {year} {2016})}\BibitemShut {NoStop}%
\bibitem [{\citenamefont {Liu}\ \emph {et~al.}(2017)\citenamefont {Liu},
  \citenamefont {Shaw}, \citenamefont {Dickey},\ and\ \citenamefont
  {Genzer}}]{liu2017sequential}%
  \BibitemOpen
  \bibfield  {author} {\bibinfo {author} {\bibfnamefont {Y.}~\bibnamefont
  {Liu}}, \bibinfo {author} {\bibfnamefont {B.}~\bibnamefont {Shaw}}, \bibinfo
  {author} {\bibfnamefont {M.~D.}\ \bibnamefont {Dickey}},\ and\ \bibinfo
  {author} {\bibfnamefont {J.}~\bibnamefont {Genzer}},\ }\bibfield  {title}
  {\bibinfo {title} {Sequential self-folding of polymer sheets},\ }\href@noop
  {} {\bibfield  {journal} {\bibinfo  {journal} {Science Advances}\ }\textbf
  {\bibinfo {volume} {3}},\ \bibinfo {pages} {e1602417} (\bibinfo {year}
  {2017})}\BibitemShut {NoStop}%
\bibitem [{\citenamefont {Lin}\ \emph {et~al.}(2020)\citenamefont {Lin},
  \citenamefont {Novelino}, \citenamefont {Wei}, \citenamefont {Alderete},
  \citenamefont {Paulino}, \citenamefont {Espinosa},\ and\ \citenamefont
  {Krishnaswamy}}]{lin2020folding}%
  \BibitemOpen
  \bibfield  {author} {\bibinfo {author} {\bibfnamefont {Z.}~\bibnamefont
  {Lin}}, \bibinfo {author} {\bibfnamefont {L.~S.}\ \bibnamefont {Novelino}},
  \bibinfo {author} {\bibfnamefont {H.}~\bibnamefont {Wei}}, \bibinfo {author}
  {\bibfnamefont {N.~A.}\ \bibnamefont {Alderete}}, \bibinfo {author}
  {\bibfnamefont {G.~H.}\ \bibnamefont {Paulino}}, \bibinfo {author}
  {\bibfnamefont {H.~D.}\ \bibnamefont {Espinosa}},\ and\ \bibinfo {author}
  {\bibfnamefont {S.}~\bibnamefont {Krishnaswamy}},\ }\bibfield  {title}
  {\bibinfo {title} {Folding at the microscale: Enabling multifunctional 3d
  origami-architected metamaterials},\ }\href@noop {} {\bibfield  {journal}
  {\bibinfo  {journal} {Small}\ }\textbf {\bibinfo {volume} {16}},\ \bibinfo
  {pages} {2002229} (\bibinfo {year} {2020})}\BibitemShut {NoStop}%
\bibitem [{\citenamefont {Miskin}\ \emph {et~al.}(2018)\citenamefont {Miskin},
  \citenamefont {Dorsey}, \citenamefont {Bircan}, \citenamefont {Han},
  \citenamefont {Muller}, \citenamefont {McEuen},\ and\ \citenamefont
  {Cohen}}]{miskin2018graphene}%
  \BibitemOpen
  \bibfield  {author} {\bibinfo {author} {\bibfnamefont {M.~Z.}\ \bibnamefont
  {Miskin}}, \bibinfo {author} {\bibfnamefont {K.~J.}\ \bibnamefont {Dorsey}},
  \bibinfo {author} {\bibfnamefont {B.}~\bibnamefont {Bircan}}, \bibinfo
  {author} {\bibfnamefont {Y.}~\bibnamefont {Han}}, \bibinfo {author}
  {\bibfnamefont {D.~A.}\ \bibnamefont {Muller}}, \bibinfo {author}
  {\bibfnamefont {P.~L.}\ \bibnamefont {McEuen}},\ and\ \bibinfo {author}
  {\bibfnamefont {I.}~\bibnamefont {Cohen}},\ }\bibfield  {title} {\bibinfo
  {title} {Graphene-based bimorphs for micron-sized, autonomous origami
  machines},\ }\href@noop {} {\bibfield  {journal} {\bibinfo  {journal}
  {Proceedings of the National Academy of Sciences}\ }\textbf {\bibinfo
  {volume} {115}},\ \bibinfo {pages} {466} (\bibinfo {year}
  {2018})}\BibitemShut {NoStop}%
\bibitem [{\citenamefont {Turner}\ \emph {et~al.}(2016)\citenamefont {Turner},
  \citenamefont {Goodwine},\ and\ \citenamefont {Sen}}]{turner2016review}%
  \BibitemOpen
  \bibfield  {author} {\bibinfo {author} {\bibfnamefont {N.}~\bibnamefont
  {Turner}}, \bibinfo {author} {\bibfnamefont {B.}~\bibnamefont {Goodwine}},\
  and\ \bibinfo {author} {\bibfnamefont {M.}~\bibnamefont {Sen}},\ }\bibfield
  {title} {\bibinfo {title} {A review of origami applications in mechanical
  engineering},\ }\href@noop {} {\bibfield  {journal} {\bibinfo  {journal}
  {Proceedings of the Institution of Mechanical Engineers, Part C: Journal of
  Mechanical Engineering Science}\ }\textbf {\bibinfo {volume} {230}},\
  \bibinfo {pages} {2345} (\bibinfo {year} {2016})}\BibitemShut {NoStop}%
\bibitem [{\citenamefont {Meloni}\ \emph {et~al.}(2021)\citenamefont {Meloni},
  \citenamefont {Cai}, \citenamefont {Zhang}, \citenamefont {Sang-Hoon~Lee},
  \citenamefont {Li}, \citenamefont {Ma}, \citenamefont {Parashkevov},\ and\
  \citenamefont {Feng}}]{meloni2021engineering}%
  \BibitemOpen
  \bibfield  {author} {\bibinfo {author} {\bibfnamefont {M.}~\bibnamefont
  {Meloni}}, \bibinfo {author} {\bibfnamefont {J.}~\bibnamefont {Cai}},
  \bibinfo {author} {\bibfnamefont {Q.}~\bibnamefont {Zhang}}, \bibinfo
  {author} {\bibfnamefont {D.}~\bibnamefont {Sang-Hoon~Lee}}, \bibinfo {author}
  {\bibfnamefont {M.}~\bibnamefont {Li}}, \bibinfo {author} {\bibfnamefont
  {R.}~\bibnamefont {Ma}}, \bibinfo {author} {\bibfnamefont {T.~E.}\
  \bibnamefont {Parashkevov}},\ and\ \bibinfo {author} {\bibfnamefont
  {J.}~\bibnamefont {Feng}},\ }\bibfield  {title} {\bibinfo {title}
  {Engineering origami: A comprehensive review of recent applications, design
  methods, and tools},\ }\href@noop {} {\bibfield  {journal} {\bibinfo
  {journal} {Advanced Science}\ ,\ \bibinfo {pages} {2000636}} (\bibinfo {year}
  {2021})}\BibitemShut {NoStop}%
\bibitem [{\citenamefont {Kim}\ \emph {et~al.}(2018)\citenamefont {Kim},
  \citenamefont {Lee}, \citenamefont {Jung},\ and\ \citenamefont
  {Cho}}]{kim2018origami}%
  \BibitemOpen
  \bibfield  {author} {\bibinfo {author} {\bibfnamefont {S.-J.}\ \bibnamefont
  {Kim}}, \bibinfo {author} {\bibfnamefont {D.-Y.}\ \bibnamefont {Lee}},
  \bibinfo {author} {\bibfnamefont {G.-P.}\ \bibnamefont {Jung}},\ and\
  \bibinfo {author} {\bibfnamefont {K.-J.}\ \bibnamefont {Cho}},\ }\bibfield
  {title} {\bibinfo {title} {An origami-inspired, self-locking robotic arm that
  can be folded flat},\ }\href@noop {} {\bibfield  {journal} {\bibinfo
  {journal} {Science Robotics}\ }\textbf {\bibinfo {volume} {3}} (\bibinfo
  {year} {2018})}\BibitemShut {NoStop}%
\bibitem [{\citenamefont {Kim}\ \emph {et~al.}(2019)\citenamefont {Kim},
  \citenamefont {Byun}, \citenamefont {Kim}, \citenamefont {Choi},
  \citenamefont {Jakobsen}, \citenamefont {Jakobsen}, \citenamefont {Lee},\
  and\ \citenamefont {Cho}}]{kim2019bioinspired}%
  \BibitemOpen
  \bibfield  {author} {\bibinfo {author} {\bibfnamefont {W.}~\bibnamefont
  {Kim}}, \bibinfo {author} {\bibfnamefont {J.}~\bibnamefont {Byun}}, \bibinfo
  {author} {\bibfnamefont {J.-K.}\ \bibnamefont {Kim}}, \bibinfo {author}
  {\bibfnamefont {W.-Y.}\ \bibnamefont {Choi}}, \bibinfo {author}
  {\bibfnamefont {K.}~\bibnamefont {Jakobsen}}, \bibinfo {author}
  {\bibfnamefont {J.}~\bibnamefont {Jakobsen}}, \bibinfo {author}
  {\bibfnamefont {D.-Y.}\ \bibnamefont {Lee}},\ and\ \bibinfo {author}
  {\bibfnamefont {K.-J.}\ \bibnamefont {Cho}},\ }\bibfield  {title} {\bibinfo
  {title} {Bioinspired dual-morphing stretchable origami},\ }\href@noop {}
  {\bibfield  {journal} {\bibinfo  {journal} {Science Robotics}\ }\textbf
  {\bibinfo {volume} {4}} (\bibinfo {year} {2019})}\BibitemShut {NoStop}%
\bibitem [{\citenamefont {Novelino}\ \emph {et~al.}(2020)\citenamefont
  {Novelino}, \citenamefont {Ze}, \citenamefont {Wu}, \citenamefont {Paulino},\
  and\ \citenamefont {Zhao}}]{novelino2020untethered}%
  \BibitemOpen
  \bibfield  {author} {\bibinfo {author} {\bibfnamefont {L.~S.}\ \bibnamefont
  {Novelino}}, \bibinfo {author} {\bibfnamefont {Q.}~\bibnamefont {Ze}},
  \bibinfo {author} {\bibfnamefont {S.}~\bibnamefont {Wu}}, \bibinfo {author}
  {\bibfnamefont {G.~H.}\ \bibnamefont {Paulino}},\ and\ \bibinfo {author}
  {\bibfnamefont {R.}~\bibnamefont {Zhao}},\ }\bibfield  {title} {\bibinfo
  {title} {Untethered control of functional origami microrobots with
  distributed actuation},\ }\href@noop {} {\bibfield  {journal} {\bibinfo
  {journal} {Proceedings of the National Academy of Sciences}\ }\textbf
  {\bibinfo {volume} {117}},\ \bibinfo {pages} {24096} (\bibinfo {year}
  {2020})}\BibitemShut {NoStop}%
\bibitem [{\citenamefont {Wu}\ \emph {et~al.}(2021)\citenamefont {Wu},
  \citenamefont {Ze}, \citenamefont {Dai}, \citenamefont {Udipi}, \citenamefont
  {Paulino},\ and\ \citenamefont {Zhao}}]{wu2021stretchable}%
  \BibitemOpen
  \bibfield  {author} {\bibinfo {author} {\bibfnamefont {S.}~\bibnamefont
  {Wu}}, \bibinfo {author} {\bibfnamefont {Q.}~\bibnamefont {Ze}}, \bibinfo
  {author} {\bibfnamefont {J.}~\bibnamefont {Dai}}, \bibinfo {author}
  {\bibfnamefont {N.}~\bibnamefont {Udipi}}, \bibinfo {author} {\bibfnamefont
  {G.~H.}\ \bibnamefont {Paulino}},\ and\ \bibinfo {author} {\bibfnamefont
  {R.}~\bibnamefont {Zhao}},\ }\bibfield  {title} {\bibinfo {title}
  {Stretchable origami robotic arm with omnidirectional bending and twisting},\
  }\href@noop {} {\bibfield  {journal} {\bibinfo  {journal} {Proceedings of the
  National Academy of Sciences}\ }\textbf {\bibinfo {volume} {118}} (\bibinfo
  {year} {2021})}\BibitemShut {NoStop}%
\bibitem [{\citenamefont {Tachi}(2009)}]{tachi2009generalization}%
  \BibitemOpen
  \bibfield  {author} {\bibinfo {author} {\bibfnamefont {T.}~\bibnamefont
  {Tachi}},\ }\bibfield  {title} {\bibinfo {title} {Generalization of
  rigid-foldable quadrilateral-mesh origami},\ }\href@noop {} {\bibfield
  {journal} {\bibinfo  {journal} {Journal of the International Association for
  Shell and Spatial Structures}\ }\textbf {\bibinfo {volume} {50}},\ \bibinfo
  {pages} {173} (\bibinfo {year} {2009})}\BibitemShut {NoStop}%
\bibitem [{\citenamefont {Waitukaitis}\ \emph {et~al.}(2015)\citenamefont
  {Waitukaitis}, \citenamefont {Menaut}, \citenamefont {Chen},\ and\
  \citenamefont {Van~Hecke}}]{waitukaitis2015origami}%
  \BibitemOpen
  \bibfield  {author} {\bibinfo {author} {\bibfnamefont {S.}~\bibnamefont
  {Waitukaitis}}, \bibinfo {author} {\bibfnamefont {R.}~\bibnamefont {Menaut}},
  \bibinfo {author} {\bibfnamefont {B.~G.-g.}\ \bibnamefont {Chen}},\ and\
  \bibinfo {author} {\bibfnamefont {M.}~\bibnamefont {Van~Hecke}},\ }\bibfield
  {title} {\bibinfo {title} {Origami multistability: From single vertices to
  metasheets},\ }\href@noop {} {\bibfield  {journal} {\bibinfo  {journal}
  {Physical review letters}\ }\textbf {\bibinfo {volume} {114}},\ \bibinfo
  {pages} {055503} (\bibinfo {year} {2015})}\BibitemShut {NoStop}%
\bibitem [{\citenamefont {Evans}\ \emph
  {et~al.}(2015{\natexlab{a}})\citenamefont {Evans}, \citenamefont {Lang},
  \citenamefont {Magleby},\ and\ \citenamefont {Howell}}]{evans2015rigidly}%
  \BibitemOpen
  \bibfield  {author} {\bibinfo {author} {\bibfnamefont {T.~A.}\ \bibnamefont
  {Evans}}, \bibinfo {author} {\bibfnamefont {R.~J.}\ \bibnamefont {Lang}},
  \bibinfo {author} {\bibfnamefont {S.~P.}\ \bibnamefont {Magleby}},\ and\
  \bibinfo {author} {\bibfnamefont {L.~L.}\ \bibnamefont {Howell}},\ }\bibfield
   {title} {\bibinfo {title} {Rigidly foldable origami gadgets and
  tessellations},\ }\href@noop {} {\bibfield  {journal} {\bibinfo  {journal}
  {Royal Society open science}\ }\textbf {\bibinfo {volume} {2}},\ \bibinfo
  {pages} {150067} (\bibinfo {year} {2015}{\natexlab{a}})}\BibitemShut
  {NoStop}%
\bibitem [{\citenamefont {Feng}\ \emph {et~al.}(2020)\citenamefont {Feng},
  \citenamefont {Dang}, \citenamefont {James},\ and\ \citenamefont
  {Plucinsky}}]{feng2020designs}%
  \BibitemOpen
  \bibfield  {author} {\bibinfo {author} {\bibfnamefont {F.}~\bibnamefont
  {Feng}}, \bibinfo {author} {\bibfnamefont {X.}~\bibnamefont {Dang}}, \bibinfo
  {author} {\bibfnamefont {R.~D.}\ \bibnamefont {James}},\ and\ \bibinfo
  {author} {\bibfnamefont {P.}~\bibnamefont {Plucinsky}},\ }\bibfield  {title}
  {\bibinfo {title} {The designs and deformations of rigidly and flat-foldable
  quadrilateral mesh origami},\ }\href@noop {} {\bibfield  {journal} {\bibinfo
  {journal} {Journal of the Mechanics and Physics of Solids}\ }\textbf
  {\bibinfo {volume} {142}},\ \bibinfo {pages} {104018} (\bibinfo {year}
  {2020})}\BibitemShut {NoStop}%
\bibitem [{\citenamefont {Dieleman}\ \emph {et~al.}(2020)\citenamefont
  {Dieleman}, \citenamefont {Vasmel}, \citenamefont {Waitukaitis},\ and\
  \citenamefont {van Hecke}}]{dieleman2020jigsaw}%
  \BibitemOpen
  \bibfield  {author} {\bibinfo {author} {\bibfnamefont {P.}~\bibnamefont
  {Dieleman}}, \bibinfo {author} {\bibfnamefont {N.}~\bibnamefont {Vasmel}},
  \bibinfo {author} {\bibfnamefont {S.}~\bibnamefont {Waitukaitis}},\ and\
  \bibinfo {author} {\bibfnamefont {M.}~\bibnamefont {van Hecke}},\ }\bibfield
  {title} {\bibinfo {title} {Jigsaw puzzle design of pluripotent origami},\
  }\href@noop {} {\bibfield  {journal} {\bibinfo  {journal} {Nature Physics}\
  }\textbf {\bibinfo {volume} {16}},\ \bibinfo {pages} {63} (\bibinfo {year}
  {2020})}\BibitemShut {NoStop}%
\bibitem [{\citenamefont {Lv}\ \emph {et~al.}(2014)\citenamefont {Lv},
  \citenamefont {Krishnaraju}, \citenamefont {Konjevod}, \citenamefont {Yu},\
  and\ \citenamefont {Jiang}}]{lv2014origami}%
  \BibitemOpen
  \bibfield  {author} {\bibinfo {author} {\bibfnamefont {C.}~\bibnamefont
  {Lv}}, \bibinfo {author} {\bibfnamefont {D.}~\bibnamefont {Krishnaraju}},
  \bibinfo {author} {\bibfnamefont {G.}~\bibnamefont {Konjevod}}, \bibinfo
  {author} {\bibfnamefont {H.}~\bibnamefont {Yu}},\ and\ \bibinfo {author}
  {\bibfnamefont {H.}~\bibnamefont {Jiang}},\ }\bibfield  {title} {\bibinfo
  {title} {Origami based mechanical metamaterials},\ }\href@noop {} {\bibfield
  {journal} {\bibinfo  {journal} {Scientific reports}\ }\textbf {\bibinfo
  {volume} {4}},\ \bibinfo {pages} {1} (\bibinfo {year} {2014})}\BibitemShut
  {NoStop}%
\bibitem [{\citenamefont {Filipov}\ \emph {et~al.}(2015)\citenamefont
  {Filipov}, \citenamefont {Tachi},\ and\ \citenamefont
  {Paulino}}]{filipov2015origami}%
  \BibitemOpen
  \bibfield  {author} {\bibinfo {author} {\bibfnamefont {E.~T.}\ \bibnamefont
  {Filipov}}, \bibinfo {author} {\bibfnamefont {T.}~\bibnamefont {Tachi}},\
  and\ \bibinfo {author} {\bibfnamefont {G.~H.}\ \bibnamefont {Paulino}},\
  }\bibfield  {title} {\bibinfo {title} {Origami tubes assembled into stiff,
  yet reconfigurable structures and metamaterials},\ }\href@noop {} {\bibfield
  {journal} {\bibinfo  {journal} {Proceedings of the National Academy of
  Sciences}\ }\textbf {\bibinfo {volume} {112}},\ \bibinfo {pages} {12321}
  (\bibinfo {year} {2015})}\BibitemShut {NoStop}%
\bibitem [{\citenamefont {Schenk}\ and\ \citenamefont
  {Guest}(2010)}]{schenk2010origami}%
  \BibitemOpen
  \bibfield  {author} {\bibinfo {author} {\bibfnamefont {M.}~\bibnamefont
  {Schenk}}\ and\ \bibinfo {author} {\bibfnamefont {S.~D.}\ \bibnamefont
  {Guest}},\ }\bibfield  {title} {\bibinfo {title} {Origami folding: A
  structural engineering approach},\ }in\ \href@noop {} {\emph {\bibinfo
  {booktitle} {5OSME, 5th international conference on Origami in Science,
  Mathematics and Education. Retrieved from: http://www. markschenk.
  com/research/\# papers}}}\ (\bibinfo {year} {2010})\BibitemShut {NoStop}%
\bibitem [{\citenamefont {Filipov}\ \emph {et~al.}(2017)\citenamefont
  {Filipov}, \citenamefont {Liu}, \citenamefont {Tachi}, \citenamefont
  {Schenk},\ and\ \citenamefont {Paulino}}]{filipov2017bar}%
  \BibitemOpen
  \bibfield  {author} {\bibinfo {author} {\bibfnamefont {E.}~\bibnamefont
  {Filipov}}, \bibinfo {author} {\bibfnamefont {K.}~\bibnamefont {Liu}},
  \bibinfo {author} {\bibfnamefont {T.}~\bibnamefont {Tachi}}, \bibinfo
  {author} {\bibfnamefont {M.}~\bibnamefont {Schenk}},\ and\ \bibinfo {author}
  {\bibfnamefont {G.~H.}\ \bibnamefont {Paulino}},\ }\bibfield  {title}
  {\bibinfo {title} {Bar and hinge models for scalable analysis of origami},\
  }\href@noop {} {\bibfield  {journal} {\bibinfo  {journal} {International
  Journal of Solids and Structures}\ }\textbf {\bibinfo {volume} {124}},\
  \bibinfo {pages} {26} (\bibinfo {year} {2017})}\BibitemShut {NoStop}%
\bibitem [{\citenamefont {Evans}\ \emph
  {et~al.}(2015{\natexlab{b}})\citenamefont {Evans}, \citenamefont
  {Silverberg},\ and\ \citenamefont {Santangelo}}]{evans2015lattice}%
  \BibitemOpen
  \bibfield  {author} {\bibinfo {author} {\bibfnamefont {A.~A.}\ \bibnamefont
  {Evans}}, \bibinfo {author} {\bibfnamefont {J.~L.}\ \bibnamefont
  {Silverberg}},\ and\ \bibinfo {author} {\bibfnamefont {C.~D.}\ \bibnamefont
  {Santangelo}},\ }\bibfield  {title} {\bibinfo {title} {Lattice mechanics of
  origami tessellations},\ }\href@noop {} {\bibfield  {journal} {\bibinfo
  {journal} {Physical Review E}\ }\textbf {\bibinfo {volume} {92}},\ \bibinfo
  {pages} {013205} (\bibinfo {year} {2015}{\natexlab{b}})}\BibitemShut
  {NoStop}%
\bibitem [{\citenamefont {Gluck}(1975)}]{gluck1975almost}%
  \BibitemOpen
  \bibfield  {author} {\bibinfo {author} {\bibfnamefont {H.}~\bibnamefont
  {Gluck}},\ }\bibfield  {title} {\bibinfo {title} {Almost all simply connected
  closed surfaces are rigid},\ }in\ \href@noop {} {\emph {\bibinfo {booktitle}
  {Geometric topology}}}\ (\bibinfo  {publisher} {Springer},\ \bibinfo {year}
  {1975})\ pp.\ \bibinfo {pages} {225--239}\BibitemShut {NoStop}%
\bibitem [{\citenamefont {Crapo}\ and\ \citenamefont
  {Whiteley}(1982)}]{crapo1982statics}%
  \BibitemOpen
  \bibfield  {author} {\bibinfo {author} {\bibfnamefont {H.}~\bibnamefont
  {Crapo}}\ and\ \bibinfo {author} {\bibfnamefont {W.}~\bibnamefont
  {Whiteley}},\ }\bibfield  {title} {\bibinfo {title} {Statics of frameworks
  and motions of panel structures: a projective geometric introduction},\
  }\href@noop {} {\bibfield  {journal} {\bibinfo  {journal} {Structural
  Topology, 1982, n{\'u}m. 6}\ } (\bibinfo {year} {1982})}\BibitemShut
  {NoStop}%
\bibitem [{\citenamefont {McInerney}\ \emph {et~al.}(2020)\citenamefont
  {McInerney}, \citenamefont {Chen}, \citenamefont {Theran}, \citenamefont
  {Santangelo},\ and\ \citenamefont {Rocklin}}]{mcinerney2020hidden}%
  \BibitemOpen
  \bibfield  {author} {\bibinfo {author} {\bibfnamefont {J.}~\bibnamefont
  {McInerney}}, \bibinfo {author} {\bibfnamefont {B.~G.-g.}\ \bibnamefont
  {Chen}}, \bibinfo {author} {\bibfnamefont {L.}~\bibnamefont {Theran}},
  \bibinfo {author} {\bibfnamefont {C.~D.}\ \bibnamefont {Santangelo}},\ and\
  \bibinfo {author} {\bibfnamefont {D.~Z.}\ \bibnamefont {Rocklin}},\
  }\bibfield  {title} {\bibinfo {title} {Hidden symmetries generate rigid
  folding mechanisms in periodic origami},\ }\href@noop {} {\bibfield
  {journal} {\bibinfo  {journal} {Proceedings of the National Academy of
  Sciences}\ }\textbf {\bibinfo {volume} {117}},\ \bibinfo {pages} {30252}
  (\bibinfo {year} {2020})}\BibitemShut {NoStop}%
\bibitem [{\citenamefont {Ting}\ and\ \citenamefont
  {Chen}(2005)}]{ting2005poisson}%
  \BibitemOpen
  \bibfield  {author} {\bibinfo {author} {\bibfnamefont {T.}~\bibnamefont
  {Ting}}\ and\ \bibinfo {author} {\bibfnamefont {T.}~\bibnamefont {Chen}},\
  }\bibfield  {title} {\bibinfo {title} {Poisson's ratio for anisotropic
  elastic materials can have no bounds},\ }\href@noop {} {\bibfield  {journal}
  {\bibinfo  {journal} {The quarterly journal of mechanics and applied
  mathematics}\ }\textbf {\bibinfo {volume} {58}},\ \bibinfo {pages} {73}
  (\bibinfo {year} {2005})}\BibitemShut {NoStop}%
\bibitem [{\citenamefont {Lovelock}\ and\ \citenamefont
  {Rund}(1989)}]{lovelock1989tensors}%
  \BibitemOpen
  \bibfield  {author} {\bibinfo {author} {\bibfnamefont {D.}~\bibnamefont
  {Lovelock}}\ and\ \bibinfo {author} {\bibfnamefont {H.}~\bibnamefont
  {Rund}},\ }\href@noop {} {\emph {\bibinfo {title} {Tensors, differential
  forms, and variational principles}}}\ (\bibinfo  {publisher} {Courier
  Corporation},\ \bibinfo {year} {1989})\BibitemShut {NoStop}%
\bibitem [{\citenamefont {Berry}\ \emph {et~al.}(2020)\citenamefont {Berry},
  \citenamefont {Lee-Trimble},\ and\ \citenamefont
  {Santangelo}}]{berry2020topological}%
  \BibitemOpen
  \bibfield  {author} {\bibinfo {author} {\bibfnamefont {M.}~\bibnamefont
  {Berry}}, \bibinfo {author} {\bibfnamefont {M.}~\bibnamefont {Lee-Trimble}},\
  and\ \bibinfo {author} {\bibfnamefont {C.}~\bibnamefont {Santangelo}},\
  }\bibfield  {title} {\bibinfo {title} {Topological transitions in the
  configuration space of non-euclidean origami},\ }\href@noop {} {\bibfield
  {journal} {\bibinfo  {journal} {Physical Review E}\ }\textbf {\bibinfo
  {volume} {101}},\ \bibinfo {pages} {043003} (\bibinfo {year}
  {2020})}\BibitemShut {NoStop}%
\bibitem [{\citenamefont {Chiang}(1984)}]{chiang1984classification}%
  \BibitemOpen
  \bibfield  {author} {\bibinfo {author} {\bibfnamefont {C.}~\bibnamefont
  {Chiang}},\ }\bibfield  {title} {\bibinfo {title} {On the classification of
  spherical four-bar linkages},\ }\href@noop {} {\bibfield  {journal} {\bibinfo
   {journal} {Mechanism and Machine Theory}\ }\textbf {\bibinfo {volume}
  {19}},\ \bibinfo {pages} {283} (\bibinfo {year} {1984})}\BibitemShut
  {NoStop}%
\bibitem [{\citenamefont {Chen}\ and\ \citenamefont
  {Santangelo}(2018)}]{chen2018branches}%
  \BibitemOpen
  \bibfield  {author} {\bibinfo {author} {\bibfnamefont {B.~G.-g.}\
  \bibnamefont {Chen}}\ and\ \bibinfo {author} {\bibfnamefont {C.~D.}\
  \bibnamefont {Santangelo}},\ }\bibfield  {title} {\bibinfo {title} {Branches
  of triangulated origami near the unfolded state},\ }\href@noop {} {\bibfield
  {journal} {\bibinfo  {journal} {Physical Review X}\ }\textbf {\bibinfo
  {volume} {8}},\ \bibinfo {pages} {011034} (\bibinfo {year}
  {2018})}\BibitemShut {NoStop}%
\bibitem [{\citenamefont {O’Rourke}(2011)}]{o2011fold}%
  \BibitemOpen
  \bibfield  {author} {\bibinfo {author} {\bibfnamefont {J.}~\bibnamefont
  {O’Rourke}},\ }\href@noop {} {\emph {\bibinfo {title} {How to fold it: the
  mathematics of linkages, origami, and polyhedra}}}\ (\bibinfo  {publisher}
  {Cambridge University Press},\ \bibinfo {year} {2011})\BibitemShut {NoStop}%
\bibitem [{\citenamefont {Klett}(2013)}]{klett2013realtime}%
  \BibitemOpen
  \bibfield  {author} {\bibinfo {author} {\bibfnamefont {Y.}~\bibnamefont
  {Klett}},\ }\bibfield  {title} {\bibinfo {title} {Realtime rigid folding
  algorithm for quadrilateral-based 1-dof tessellations},\ }in\ \href@noop {}
  {\emph {\bibinfo {booktitle} {International Design Engineering Technical
  Conferences and Computers and Information in Engineering Conference}}},\
  Vol.\ \bibinfo {volume} {55942}\ (\bibinfo {organization} {American Society
  of Mechanical Engineers},\ \bibinfo {year} {2013})\ p.\ \bibinfo {pages}
  {V06BT07A031}\BibitemShut {NoStop}%
\bibitem [{\citenamefont {Zhou}\ \emph {et~al.}(2016)\citenamefont {Zhou},
  \citenamefont {Zang},\ and\ \citenamefont {You}}]{zhou2016origami}%
  \BibitemOpen
  \bibfield  {author} {\bibinfo {author} {\bibfnamefont {X.}~\bibnamefont
  {Zhou}}, \bibinfo {author} {\bibfnamefont {S.}~\bibnamefont {Zang}},\ and\
  \bibinfo {author} {\bibfnamefont {Z.}~\bibnamefont {You}},\ }\bibfield
  {title} {\bibinfo {title} {Origami mechanical metamaterials based on the
  miura-derivative fold patterns},\ }\href@noop {} {\bibfield  {journal}
  {\bibinfo  {journal} {Proceedings of the Royal Society A: Mathematical,
  Physical and Engineering Sciences}\ }\textbf {\bibinfo {volume} {472}},\
  \bibinfo {pages} {20160361} (\bibinfo {year} {2016})}\BibitemShut {NoStop}%
\bibitem [{\citenamefont {Overvelde}\ \emph {et~al.}(2016)\citenamefont
  {Overvelde}, \citenamefont {De~Jong}, \citenamefont {Shevchenko},
  \citenamefont {Becerra}, \citenamefont {Whitesides}, \citenamefont {Weaver},
  \citenamefont {Hoberman},\ and\ \citenamefont
  {Bertoldi}}]{overvelde2016three}%
  \BibitemOpen
  \bibfield  {author} {\bibinfo {author} {\bibfnamefont {J.~T.}\ \bibnamefont
  {Overvelde}}, \bibinfo {author} {\bibfnamefont {T.~A.}\ \bibnamefont
  {De~Jong}}, \bibinfo {author} {\bibfnamefont {Y.}~\bibnamefont {Shevchenko}},
  \bibinfo {author} {\bibfnamefont {S.~A.}\ \bibnamefont {Becerra}}, \bibinfo
  {author} {\bibfnamefont {G.~M.}\ \bibnamefont {Whitesides}}, \bibinfo
  {author} {\bibfnamefont {J.~C.}\ \bibnamefont {Weaver}}, \bibinfo {author}
  {\bibfnamefont {C.}~\bibnamefont {Hoberman}},\ and\ \bibinfo {author}
  {\bibfnamefont {K.}~\bibnamefont {Bertoldi}},\ }\bibfield  {title} {\bibinfo
  {title} {A three-dimensional actuated origami-inspired transformable
  metamaterial with multiple degrees of freedom},\ }\href@noop {} {\bibfield
  {journal} {\bibinfo  {journal} {Nature communications}\ }\textbf {\bibinfo
  {volume} {7}},\ \bibinfo {pages} {1} (\bibinfo {year} {2016})}\BibitemShut
  {NoStop}%
\bibitem [{\citenamefont {Liu}\ and\ \citenamefont
  {Paulino}(2017)}]{liu2017nonlinear}%
  \BibitemOpen
  \bibfield  {author} {\bibinfo {author} {\bibfnamefont {K.}~\bibnamefont
  {Liu}}\ and\ \bibinfo {author} {\bibfnamefont {G.}~\bibnamefont {Paulino}},\
  }\bibfield  {title} {\bibinfo {title} {Nonlinear mechanics of non-rigid
  origami: an efficient computational approach},\ }\href@noop {} {\bibfield
  {journal} {\bibinfo  {journal} {Proceedings of the Royal Society A:
  Mathematical, Physical and Engineering Sciences}\ }\textbf {\bibinfo {volume}
  {473}},\ \bibinfo {pages} {20170348} (\bibinfo {year} {2017})}\BibitemShut
  {NoStop}%
\bibitem [{\citenamefont {Pinson}\ \emph {et~al.}(2017)\citenamefont {Pinson},
  \citenamefont {Stern}, \citenamefont {Ferrero}, \citenamefont {Witten},
  \citenamefont {Chen},\ and\ \citenamefont {Murugan}}]{pinson2017self}%
  \BibitemOpen
  \bibfield  {author} {\bibinfo {author} {\bibfnamefont {M.~B.}\ \bibnamefont
  {Pinson}}, \bibinfo {author} {\bibfnamefont {M.}~\bibnamefont {Stern}},
  \bibinfo {author} {\bibfnamefont {A.~C.}\ \bibnamefont {Ferrero}}, \bibinfo
  {author} {\bibfnamefont {T.~A.}\ \bibnamefont {Witten}}, \bibinfo {author}
  {\bibfnamefont {E.}~\bibnamefont {Chen}},\ and\ \bibinfo {author}
  {\bibfnamefont {A.}~\bibnamefont {Murugan}},\ }\bibfield  {title} {\bibinfo
  {title} {Self-folding origami at any energy scale},\ }\href@noop {}
  {\bibfield  {journal} {\bibinfo  {journal} {Nature communications}\ }\textbf
  {\bibinfo {volume} {8}},\ \bibinfo {pages} {1} (\bibinfo {year}
  {2017})}\BibitemShut {NoStop}%
\bibitem [{\citenamefont {Liu}\ \emph {et~al.}(2020)\citenamefont {Liu},
  \citenamefont {Novelino}, \citenamefont {Gardoni},\ and\ \citenamefont
  {Paulino}}]{liu2020big}%
  \BibitemOpen
  \bibfield  {author} {\bibinfo {author} {\bibfnamefont {K.}~\bibnamefont
  {Liu}}, \bibinfo {author} {\bibfnamefont {L.~S.}\ \bibnamefont {Novelino}},
  \bibinfo {author} {\bibfnamefont {P.}~\bibnamefont {Gardoni}},\ and\ \bibinfo
  {author} {\bibfnamefont {G.~H.}\ \bibnamefont {Paulino}},\ }\bibfield
  {title} {\bibinfo {title} {Big influence of small random imperfections in
  origami-based metamaterials},\ }\href@noop {} {\bibfield  {journal} {\bibinfo
   {journal} {Proceedings of the Royal Society A}\ }\textbf {\bibinfo {volume}
  {476}},\ \bibinfo {pages} {20200236} (\bibinfo {year} {2020})}\BibitemShut
  {NoStop}%
\bibitem [{\citenamefont {Stern}\ \emph {et~al.}(2017)\citenamefont {Stern},
  \citenamefont {Pinson},\ and\ \citenamefont {Murugan}}]{stern2017complexity}%
  \BibitemOpen
  \bibfield  {author} {\bibinfo {author} {\bibfnamefont {M.}~\bibnamefont
  {Stern}}, \bibinfo {author} {\bibfnamefont {M.~B.}\ \bibnamefont {Pinson}},\
  and\ \bibinfo {author} {\bibfnamefont {A.}~\bibnamefont {Murugan}},\
  }\bibfield  {title} {\bibinfo {title} {The complexity of folding self-folding
  origami},\ }\href@noop {} {\bibfield  {journal} {\bibinfo  {journal}
  {Physical Review X}\ }\textbf {\bibinfo {volume} {7}},\ \bibinfo {pages}
  {041070} (\bibinfo {year} {2017})}\BibitemShut {NoStop}%
\bibitem [{\citenamefont {Grey}\ \emph {et~al.}(2018)\citenamefont {Grey},
  \citenamefont {Schenk},\ and\ \citenamefont {Scarpa}}]{grey2018local}%
  \BibitemOpen
  \bibfield  {author} {\bibinfo {author} {\bibfnamefont {S.}~\bibnamefont
  {Grey}}, \bibinfo {author} {\bibfnamefont {M.}~\bibnamefont {Schenk}},\ and\
  \bibinfo {author} {\bibfnamefont {F.}~\bibnamefont {Scarpa}},\ }\bibfield
  {title} {\bibinfo {title} {Local actuation of tubular origami},\ }in\
  \href@noop {} {\emph {\bibinfo {booktitle} {Proceedings of the seventh
  meeting of Origami, Science, Mathematics and Education, Oxford, UK}}}\
  (\bibinfo {year} {2018})\ pp.\ \bibinfo {pages} {4--7}\BibitemShut {NoStop}%
\bibitem [{\citenamefont {Grey}\ \emph {et~al.}(2019)\citenamefont {Grey},
  \citenamefont {Scarpa},\ and\ \citenamefont {Schenk}}]{grey2019strain}%
  \BibitemOpen
  \bibfield  {author} {\bibinfo {author} {\bibfnamefont {S.~W.}\ \bibnamefont
  {Grey}}, \bibinfo {author} {\bibfnamefont {F.}~\bibnamefont {Scarpa}},\ and\
  \bibinfo {author} {\bibfnamefont {M.}~\bibnamefont {Schenk}},\ }\bibfield
  {title} {\bibinfo {title} {Strain reversal in actuated origami structures},\
  }\href@noop {} {\bibfield  {journal} {\bibinfo  {journal} {Physical review
  letters}\ }\textbf {\bibinfo {volume} {123}},\ \bibinfo {pages} {025501}
  (\bibinfo {year} {2019})}\BibitemShut {NoStop}%
\bibitem [{\citenamefont {Thiria}\ and\ \citenamefont
  {Adda-Bedia}(2011)}]{thiria2011relaxation}%
  \BibitemOpen
  \bibfield  {author} {\bibinfo {author} {\bibfnamefont {B.}~\bibnamefont
  {Thiria}}\ and\ \bibinfo {author} {\bibfnamefont {M.}~\bibnamefont
  {Adda-Bedia}},\ }\bibfield  {title} {\bibinfo {title} {Relaxation mechanisms
  in the unfolding of thin sheets},\ }\href@noop {} {\bibfield  {journal}
  {\bibinfo  {journal} {Physical review letters}\ }\textbf {\bibinfo {volume}
  {107}},\ \bibinfo {pages} {025506} (\bibinfo {year} {2011})}\BibitemShut
  {NoStop}%
\bibitem [{\citenamefont {Jules}\ \emph {et~al.}(2020)\citenamefont {Jules},
  \citenamefont {Lechenault},\ and\ \citenamefont
  {Adda-Bedia}}]{jules2020plasticity}%
  \BibitemOpen
  \bibfield  {author} {\bibinfo {author} {\bibfnamefont {T.}~\bibnamefont
  {Jules}}, \bibinfo {author} {\bibfnamefont {F.}~\bibnamefont {Lechenault}},\
  and\ \bibinfo {author} {\bibfnamefont {M.}~\bibnamefont {Adda-Bedia}},\
  }\bibfield  {title} {\bibinfo {title} {Plasticity and aging of folded elastic
  sheets},\ }\href@noop {} {\bibfield  {journal} {\bibinfo  {journal} {Physical
  Review E}\ }\textbf {\bibinfo {volume} {102}},\ \bibinfo {pages} {033005}
  (\bibinfo {year} {2020})}\BibitemShut {NoStop}%
\bibitem [{\citenamefont {Zadpoor}(2016)}]{zadpoor2016mechanical}%
  \BibitemOpen
  \bibfield  {author} {\bibinfo {author} {\bibfnamefont {A.~A.}\ \bibnamefont
  {Zadpoor}},\ }\bibfield  {title} {\bibinfo {title} {Mechanical
  meta-materials},\ }\href@noop {} {\bibfield  {journal} {\bibinfo  {journal}
  {Materials Horizons}\ }\textbf {\bibinfo {volume} {3}},\ \bibinfo {pages}
  {371} (\bibinfo {year} {2016})}\BibitemShut {NoStop}%
\bibitem [{\citenamefont {Bertoldi}\ \emph {et~al.}(2017)\citenamefont
  {Bertoldi}, \citenamefont {Vitelli}, \citenamefont {Christensen},\ and\
  \citenamefont {Van~Hecke}}]{bertoldi2017flexible}%
  \BibitemOpen
  \bibfield  {author} {\bibinfo {author} {\bibfnamefont {K.}~\bibnamefont
  {Bertoldi}}, \bibinfo {author} {\bibfnamefont {V.}~\bibnamefont {Vitelli}},
  \bibinfo {author} {\bibfnamefont {J.}~\bibnamefont {Christensen}},\ and\
  \bibinfo {author} {\bibfnamefont {M.}~\bibnamefont {Van~Hecke}},\ }\bibfield
  {title} {\bibinfo {title} {Flexible mechanical metamaterials},\ }\href@noop
  {} {\bibfield  {journal} {\bibinfo  {journal} {Nature Reviews Materials}\
  }\textbf {\bibinfo {volume} {2}},\ \bibinfo {pages} {1} (\bibinfo {year}
  {2017})}\BibitemShut {NoStop}%
\bibitem [{\citenamefont {Yu}\ \emph {et~al.}(2018)\citenamefont {Yu},
  \citenamefont {Zhou}, \citenamefont {Liang}, \citenamefont {Jiang},\ and\
  \citenamefont {Wu}}]{yu2018mechanical}%
  \BibitemOpen
  \bibfield  {author} {\bibinfo {author} {\bibfnamefont {X.}~\bibnamefont
  {Yu}}, \bibinfo {author} {\bibfnamefont {J.}~\bibnamefont {Zhou}}, \bibinfo
  {author} {\bibfnamefont {H.}~\bibnamefont {Liang}}, \bibinfo {author}
  {\bibfnamefont {Z.}~\bibnamefont {Jiang}},\ and\ \bibinfo {author}
  {\bibfnamefont {L.}~\bibnamefont {Wu}},\ }\bibfield  {title} {\bibinfo
  {title} {Mechanical metamaterials associated with stiffness, rigidity and
  compressibility: A brief review},\ }\href@noop {} {\bibfield  {journal}
  {\bibinfo  {journal} {Progress in Materials Science}\ }\textbf {\bibinfo
  {volume} {94}},\ \bibinfo {pages} {114} (\bibinfo {year} {2018})}\BibitemShut
  {NoStop}%
\bibitem [{\citenamefont {Barchiesi}\ \emph {et~al.}(2019)\citenamefont
  {Barchiesi}, \citenamefont {Spagnuolo},\ and\ \citenamefont
  {Placidi}}]{barchiesi2019mechanical}%
  \BibitemOpen
  \bibfield  {author} {\bibinfo {author} {\bibfnamefont {E.}~\bibnamefont
  {Barchiesi}}, \bibinfo {author} {\bibfnamefont {M.}~\bibnamefont
  {Spagnuolo}},\ and\ \bibinfo {author} {\bibfnamefont {L.}~\bibnamefont
  {Placidi}},\ }\bibfield  {title} {\bibinfo {title} {Mechanical metamaterials:
  a state of the art},\ }\href@noop {} {\bibfield  {journal} {\bibinfo
  {journal} {Mathematics and Mechanics of Solids}\ }\textbf {\bibinfo {volume}
  {24}},\ \bibinfo {pages} {212} (\bibinfo {year} {2019})}\BibitemShut
  {NoStop}%
\bibitem [{\citenamefont {Surjadi}\ \emph {et~al.}(2019)\citenamefont
  {Surjadi}, \citenamefont {Gao}, \citenamefont {Du}, \citenamefont {Li},
  \citenamefont {Xiong}, \citenamefont {Fang},\ and\ \citenamefont
  {Lu}}]{surjadi2019mechanical}%
  \BibitemOpen
  \bibfield  {author} {\bibinfo {author} {\bibfnamefont {J.~U.}\ \bibnamefont
  {Surjadi}}, \bibinfo {author} {\bibfnamefont {L.}~\bibnamefont {Gao}},
  \bibinfo {author} {\bibfnamefont {H.}~\bibnamefont {Du}}, \bibinfo {author}
  {\bibfnamefont {X.}~\bibnamefont {Li}}, \bibinfo {author} {\bibfnamefont
  {X.}~\bibnamefont {Xiong}}, \bibinfo {author} {\bibfnamefont {N.~X.}\
  \bibnamefont {Fang}},\ and\ \bibinfo {author} {\bibfnamefont
  {Y.}~\bibnamefont {Lu}},\ }\bibfield  {title} {\bibinfo {title} {Mechanical
  metamaterials and their engineering applications},\ }\href@noop {} {\bibfield
   {journal} {\bibinfo  {journal} {Advanced Engineering Materials}\ }\textbf
  {\bibinfo {volume} {21}},\ \bibinfo {pages} {1800864} (\bibinfo {year}
  {2019})}\BibitemShut {NoStop}%
\bibitem [{\citenamefont {Wegener}(2013)}]{wegener2013metamaterials}%
  \BibitemOpen
  \bibfield  {author} {\bibinfo {author} {\bibfnamefont {M.}~\bibnamefont
  {Wegener}},\ }\bibfield  {title} {\bibinfo {title} {Metamaterials beyond
  optics},\ }\href@noop {} {\bibfield  {journal} {\bibinfo  {journal}
  {Science}\ }\textbf {\bibinfo {volume} {342}},\ \bibinfo {pages} {939}
  (\bibinfo {year} {2013})}\BibitemShut {NoStop}%
\bibitem [{\citenamefont {Kadic}\ \emph {et~al.}(2013)\citenamefont {Kadic},
  \citenamefont {B{\"u}ckmann}, \citenamefont {Schittny},\ and\ \citenamefont
  {Wegener}}]{kadic2013metamaterials}%
  \BibitemOpen
  \bibfield  {author} {\bibinfo {author} {\bibfnamefont {M.}~\bibnamefont
  {Kadic}}, \bibinfo {author} {\bibfnamefont {T.}~\bibnamefont {B{\"u}ckmann}},
  \bibinfo {author} {\bibfnamefont {R.}~\bibnamefont {Schittny}},\ and\
  \bibinfo {author} {\bibfnamefont {M.}~\bibnamefont {Wegener}},\ }\bibfield
  {title} {\bibinfo {title} {Metamaterials beyond electromagnetism},\
  }\href@noop {} {\bibfield  {journal} {\bibinfo  {journal} {Reports on
  Progress in physics}\ }\textbf {\bibinfo {volume} {76}},\ \bibinfo {pages}
  {126501} (\bibinfo {year} {2013})}\BibitemShut {NoStop}%
\end{thebibliography}%

\newpage
\onecolumngrid
\appendix
\newcommand{\embedding}{\mathbf{X}}
\newcommand{\firstform}{I}
\newcommand{\secondform}{II}
\newcommand{\tangentvector}{\hat{t}}
\newcommand{\cellrotation}{\mathbf{S}}
\newcommand{\interior}{\sigma}
\newcolumntype{M}{>{$}c<{$}}

\section{Rigidly foldable ground states of four-parallelogram origami} \label{appendix:system}

Here, we provide additional details about the four-parallelogram origami family of crease patterns. First, we discuss subsets and the limiting cases of previously studied patterns. Then, we parameterize the degenerate ground states of generic four-parallelogram origami. For completeness, we discuss this parameterization in regards to branching from the flattened state of developable crease patterns.

\subsection{Four-parallelogram origami and limiting cases}

\begin{figure}
    \centering
    \includegraphics{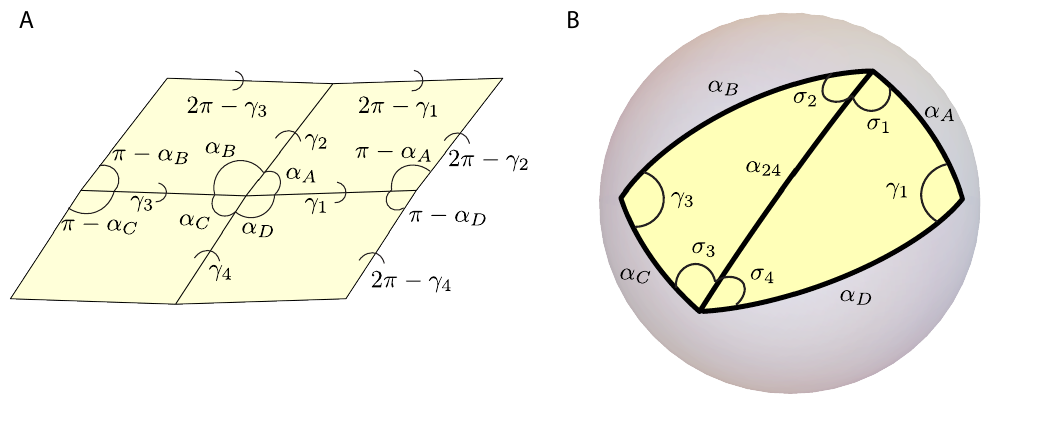}
    \caption{(A) The unit cell of four-parallelogram origami is characterized by the four sector angles, $\sectorangle_A, \sectorangle_B, \sectorangle_C$, and $\sectorangle_D$ which are identical at non-adjacent corners and supplementary, $\pi - \sectorangle$, in adjacent corners for parallelogram faces. The configuration of such a geometry is specified by the four dihedral angles, $\dihedralangle_1, \dihedralangle_2, \dihedralangle_3,$ and $\dihedralangle_4$ which are complementary, $2\pi - \dihedralangle$, on parallel edges to maintain spatial periodicity. (B) The projection of the central vertex in panel (A) onto the unit cell yields a spherical quadrilateral, whose edges have arc lengths subtending the sector angles and interior angles subtending the dihedral angles, that is triangulated via the great circle of arc length $\sectorangle_{24}$ which divides the interior angles $\dihedralangle_2 = \interior_1 + \interior_2$ and $\dihedralangle_4 = \interior_3 + \interior_4$.}
    \label{fig:geometry}
\end{figure}

Four-parallelogram origami geometries can be specified by four sector angles and four edge lengths, both defined in the vicinity of a single vertex as shown in \ref{fig:geometry}A. Since each face is a parallelogram, the adjacent sector angles are supplementary, $\pi - \sectorangle$ and the non-adjacent angles are identical. Furthermore, since there are only four parallelogram faces, the four edge lengths determine all eight of the edges in the unit cell. Hence, generic members correspond to a point in the eight-dimensional space of geometries where one such dimension simply rescales the entire sheet. 

This eight-dimensional space of geometries contains multiple subspaces of interest. First, developable crease patterns have the one-dimensional constraint on their four sector angles:

\begin{equation} \label{eq:developability}
\sectorangle_A + \sectorangle_B + \sectorangle_C + \sectorangle_D = 0,
\end{equation}

\noindent
which is a seven-dimensional subspace. Similarly, orthotropic crease patterns have a four-dimensional constraint that couples their sector angles and edge directions:

\begin{equation}\label{eq:orthotropic}
\latticevector_1 \cdot \latticevector_2 = (\globaledgevector_1 + \globaledgevector_3) \cdot (\globaledgevector_2 + \globaledgevector_4) = \globaledge_1 \globaledge_2 \sectorangle_A - \globaledge_2 \globaledge_3 \sectorangle_B + \globaledge_3 \globaledge_4 \sectorangle_C -\globaledge_1 \globaledge_4 \sectorangle_D = 0,
\end{equation}

\noindent
which again is a seven-dimensional subspace. The Miura-ori belongs to the special case of developable, orthotropic crease patterns satisfying $\sectorangle_A = \sectorangle_B$, $\sectorangle_C = \sectorangle_D = \pi - \sectorangle_A$~\cite{schenk2013geometry,wei2013geometric}. The eggbox belongs to the special case of orthotropic crease patterns satisfying $\sectorangle_A = \sectorangle_B = \sectorangle_C = \sectorangle_D$~\cite{nassar2017curvature}. Both of these crease patterns are special cases of the orthotropic Morph which itself satisfies $\sectorangle_A = \sectorangle_B$, $\sectorangle_C = \sectorangle_D$~\cite{pratapa2019geometric}.

\subsection{Degenerate ground states}

The ground states of these crease patterns are parameterized by the four dihedral angles defined in the vicinity of a vertex as shown in Fig.~\ref{fig:geometry}A. Since the crease pattern is spatially periodic, the remaining dihedral angles must be the complement of their parallel counterparts, $2\pi - \dihedralangle$. Hence, the unit cell can be rigidly folded provided that the changes to these four dihedral angles are compatible with the four sector angles. Such configurations can be provided by applying spherical trigonometry to the projection of the four-coordinated vertex onto the unit sphere as shown in Fig.~\ref{fig:geometry}B. 

The configurations of such a spherical quadrilateral are determined by triangulating the quadrilateral with a single great circle and enforcing compatibility between the two resulting spherical triangles shown in Fig.~\ref{fig:geometry}B. These triangles obey the spherical trigonometric relations:

\begin{align}
\begin{split}
\cos \sectorangle_{24} = \cos \sectorangle_A \cos \sectorangle_D + \sin \sectorangle_A \sin \sectorangle_D \cos \dihedralangle_1
& = \cos \sectorangle_B \cos \sectorangle_C + \sin \sectorangle_B \sin \sectorangle_C \cos \dihedralangle_3, \\
\cos \interior_1 = \frac{ \cos \sectorangle_D - \cos \sectorangle_{24} \cos \sectorangle_A }{ \sin \sectorangle_{24} \sin \sectorangle_A },
\hspace{20pt} &
\cos \interior_4 = \frac{ \cos \sectorangle_A - \cos \sectorangle_{24} \cos \sectorangle_D }{ \sin \sectorangle_{24} \sin \sectorangle_D },
\\
\cos \interior_2 = \frac{ \cos \sectorangle_C - \cos \sectorangle_{24} \cos \sectorangle_B }{ \sin \sectorangle_{24} \sin \sectorangle_C },
\hspace{20pt} &
\cos \interior_3 = \frac{ \cos \sectorangle_B - \cos \sectorangle_{24} \cos \sectorangle_C }{ \sin \sectorangle_{24} \sin \sectorangle_B },
\end{split} \\
\frac{ \sin \dihedralangle_1 }{ \sin \sectorangle_{24} } = 
\frac{ \sin \interior_1}{ \sin \sectorangle_D} = 
\frac{ \sin \interior_4}{ \sin \sectorangle_A},
\hspace{20pt} &
\frac{ \sin \dihedralangle_3 }{ \sin \sectorangle_{24} } = 
\frac{ \sin \interior_2}{ \sin \sectorangle_C} = 
\frac{ \sin \interior_3}{ \sin \sectorangle_B}.
\end{align}

\noindent
These formulae give $\dihedralangle_3$ in terms of $\dihedralangle_1$ or vice versa noting there are always two solutions because $\arccos$ is multivalued over the unit circle. Once this is chosen, the diagonal $\sectorangle_{24}$ can be determined to compute the remaining interior angles $\interior_i$ and sum them for the last two dihedral angles $\dihedralangle_2, \dihedralangle_4$. Importantly, the $\arctan$ function ensures that the branches are appropriately determined:

\begin{align}
\dihedralangle_2 = \arctan \frac{ \sin \dihedralangle_1 \sin \sectorangle_D \sin \sectorangle_A }{ \cos \sectorangle_D - \cos \sectorangle_{24} \cos \sectorangle_A } 
& + \arctan \frac{ \sin \dihedralangle_1 \sin \sectorangle_A \sin \sectorangle_D }{ \cos \sectorangle_A - \cos \sectorangle_{24} \cos \sectorangle_D } \\
\dihedralangle_4 = \arctan \frac{ \sin \dihedralangle_3 \sin \sectorangle_C \sin \sectorangle_B}{ \cos \sectorangle_C - \cos \sectorangle_{24} \cos \sectorangle_B } 
& + \arctan \frac{ \sin \dihedralangle_3 \sin \sectorangle_B \sin \sectorangle_C}{ \cos \sectorangle_B - \cos \sectorangle_{24} \cos \sectorangle_C }.
\end{align}

\noindent
Once these dihedral angles are determined at a single vertex, it is easy to see the compatibility of adjacent vertices in  four-parallelogram origami by substitution of the appropriate sector angles: when the dihedral angles are fixed to be identical on an edge shared by two vertices, the supplementary condition on the sector angles ensures that the edges which are not shared have complementary dihedral angles. 

In principle, the choice to complete this parameterization by varying $\dihedralangle_1$ or $\dihedralangle_3$ is trivial; however, their domains are generically distinct because the domain of $\arccos$ used to compute the diagonal restricts the admissible dihedral angles. For example:

\begin{align}
\dihedralangle_1 = 0 \implies & \sectorangle_{24} = \arccos \Big( \cos \sectorangle_A \cos \sectorangle_D + \sin \sectorangle_A \sin \sectorangle_D \Big), \\
\dihedralangle_1 = \pi \implies & \sectorangle_{24} = \arccos \Big( \cos \sectorangle_A \cos \sectorangle_D - \sin \sectorangle_A \sin \sectorangle_D \Big),
\end{align}

\noindent
which must hold for $\dihedralangle_3$ replacing $\sectorangle_A \rightarrow \sectorangle_B$ and $\sectorangle_D \rightarrow \sectorangle_C$ though $\sectorangle_{24}$ is only real-valued provided the arguments are over the interval $[0,1]$ implying one edge may open or close while the other may be locked from doing so. Accordingly, the diagonal is bounded:

\begin{equation}
\sectorangle_{24}^{\text{min}} = \max \Big( \lvert \sectorangle_A - \sectorangle_D \rvert, \lvert \sectorangle_B - \sectorangle_C \rvert \Big),
\hspace{20pt}
\sectorangle_{24}^{\text{max}} = \min \Big( \lvert \sectorangle_A + \sectorangle_D \rvert, \lvert \sectorangle_B + \sectorangle_C \rvert \Big),
\end{equation}

\noindent
where the inner product $\lvert x \rvert = \min ( x, 2 \pi - x )$ is the geodesic length of the corresponding great circle taking $x$ to always be positive valued. Hence, there are two distinct cases: the same sector angle pair determines both bounds or each sector angle pair determines a single bound. In the former, the corresponding dihedral angle's domain contains both the opened state $\pi$ and the closed state $0$ (and hence also $2\pi$ implying the configuration space is a non-contractible loop. In the latter, the corresponding dihedral angle's domain contains only contains $0$ or $\pi$ so that these two solutions join to form a single contractible loop in the configuration space. In special cases, both pairs simultaneously bound the diagonal so that the vertex is flat-foldable (in the generalized sense that two of the four dihedral angles may still be $\pi$) or developable.

\subsection{Folding near the flattened state}

Generically, the diagonal satisfies

\begin{equation}
\cos \sectorangle_{24} = \cos \sectorangle_A \cos \sectorangle_D + \sin \sectorangle_A \sin \sectorangle_D \cos \dihedralangle_1,
\end{equation}

\noindent
which, in the flattened state of a developable crease pattern, $\dihedralangle_1 = 0$, take on the value:

\begin{equation}
\sectorangle_{24}^f = \min \Big( \sectorangle_A + \sectorangle_D, 2 \pi - (\sectorangle_A + \sectorangle_D) \Big).
\end{equation}

\noindent
Moreover the developability condition $\sum_i \sectorangle_{A_i} = 2 \pi$ indicates that the dividing great circle lies either along $\sectorangle_A + \sectorangle_D$ or $\sectorangle_B + \sectorangle_C$. This means if $\interior_1, \interior_2$ are the two interior angles obtained by the great circle $\sectorangle_{24}^f$ dividing $\dihedralangle_2$ then the $\dihedralangle_2 = \interior_1$ or $\interior_2$ where the alternate interior angle is zero. Suppose that this is $\interior_1$ so that $\sectorangle_{24}^f = \sectorangle_B + \sectorangle_C$ where the following holds for $\interior_2$ under the substitution $\dihedralangle_1 \rightarrow \dihedralangle_3$, $\sectorangle_A \rightarrow \sectorangle_B$, and $\sectorangle_D \rightarrow \sectorangle_C$.  Then by the spherical law of sines and cosines:

\begin{equation}
\frac{ \sin \interior_1 }{ \sin \sectorangle_D } = \frac{ \sin \dihedralangle_1 }{ \sin \sectorangle_{24}^f },
\hspace{20pt}
\cos \sectorangle_D = \cos \sectorangle_A \cos \sectorangle_{24}^f - \sin \sectorangle_A \sin \sectorangle_{24}^f \cos \interior_1,
\hspace{20pt}
\tan \interior_1 = \frac{ \sin \dihedralangle_1 \sin \sectorangle_A \sin \sectorangle_D }{ \cos \sectorangle_B - \cos \sectorangle_A \cos \sectorangle_{24}^f }.
\end{equation}

\noindent 
On the other hand, $\sectorangle_{24}^f$ divides when the law of cosines satisfies:

\begin{equation}
\cos \sectorangle_{24}^f = \cos \sectorangle_B \cos \sectorangle_C + \sin \sectorangle_B \sin \sectorangle_C \cos \dihedralangle_3.
\end{equation}

\noindent
Thus, expanding about $\interior_1 = \pi$ and $\dihedralangle_1 = \pi$ yields the linearly compatible differentials

\begin{align}
\begin{split}
d \dihedralangle_2 & = - \frac{d \dihedralangle_1 \sin \sectorangle_A \sin \sectorangle_D }{ \cos \sectorangle_D - \cos \sectorangle_A \cos \sectorangle_{24}^f },
\hspace{20pt} 
d \dihedralangle_4 = - \frac{d \dihedralangle_1 \sin \sectorangle_A \sin \sectorangle_D }{ \cos \sectorangle_A - \cos \sectorangle_D \cos \sectorangle_{24}^f }, \\
d \dihedralangle_3 & = \pm \Big( \frac{ \sin \sectorangle_A \sin \sectorangle_D (d \dihedralangle_1)^2 + \cos \sectorangle_B \cos \sectorangle_C - \cos \sectorangle_A \cos \sectorangle_D }{ \sin \sectorangle_B \sin \sectorangle_C } \Big)^{\frac{1}{2}},
\end{split}
\end{align}

\noindent
which determine the linear planar modes that generate the two branches intersecting at the flattened state.

\section{Compatibility conditions in four-parallelogram origami} \label{appendix:compatibility}
Here, we explicitly derive the linear compatibility conditions and their solutions for four-parallelogram origami. We also compute the non-trivial face amplitudes for the antisymmetric bend mode and discuss normalization of the linear isometries.

\begin{table}
\begin{tabular}{| M | M | M | M | M |}
\hline
\frac{\foldcoefficient_i}{\localedge_i} & a & b & c & d \\
\hline
1 & \globalcoefficient_1 & - \globalcoefficient_3 & \globalcoefficient_3 & - \globalcoefficient_1 \\
\hline
2 & \globalcoefficient_2 & - \globalcoefficient_2 & \globalcoefficient_4 & -\globalcoefficient_4 \\
\hline
3 & -\globalcoefficient_3 & \globalcoefficient_1 & - \globalcoefficient_1 & \globalcoefficient_3 \\
\hline
4 & -\globalcoefficient_4 & \globalcoefficient_4 & -\globalcoefficient_2 & \globalcoefficient_2 \\
\hline
\end{tabular}
\caption{The local folding coefficients written in terms of the global folding coefficients.}
\label{table:coefficients}
\end{table}

\subsection{Four-parallelogram compatibility matrix}

The linear isometries of any parallelogram-based origami sheet are spanned by the vertex amplitudes satisfying the vertex compatibility condition

\begin{equation}
    \sum_{i'} \Big( \frac{\foldcoefficient_{i'}^a}{\localedge_{i'}^a} \vertexamplitude^a - \frac{\foldcoefficient_{i'+2}^{a'}}{\localedge_{i'+2}^{a'}} \vertexamplitude^{a'} \Big) = 0,
\end{equation}

\noindent
in addition to the uniform face-bending mode where $\faceamplitude = 1$ on every face. In four-parallelogram origami, these \emph{local} coefficients are proportional to the \emph{global} coefficients:

\begin{equation}
\globalcoefficient_i \equiv \frac{\globaledgevector_{i+2} \cdot \globaledgevector_{i+3} \times \globaledgevector_{i+4}}{\edgeproduct} = \frac{\globaledgedirection_{i+2} \cdot \globaledgedirection_{i+3} \times \globaledgedirection_{i+4} }{ \globaledge_i}, \quad
\edgeproduct \equiv \globaledge_1 \globaledge_2 \globaledge_3 \globaledge_4,
\end{equation}

\noindent 
as shown in Table~\ref{table:coefficients}. Denoting the sums and differences $\globalcoefficient_{ij}^{\pm} = \globalcoefficient_i \pm \globalcoefficient_j$, the corresponding compatibility matrix is:

\begin{equation}
\compatibility = 
\begin{pmatrix}
\globalcoefficient_{13}^- + \globalcoefficient_{24}^- & - \globalcoefficient_{13}^- & 0 & - \globalcoefficient_{24}^- \\
-\globalcoefficient_{13}^- & \globalcoefficient_{13}^- - \globalcoefficient_{24}^- & \globalcoefficient_{24}^- & 0 \\
0 & \globalcoefficient_{24}^- & - \globalcoefficient_{13}^- - \globalcoefficient_{24}^- & \globalcoefficient_{13}^- \\
- \globalcoefficient_{24}^- & 0 & \globalcoefficient_{13}^- & -\globalcoefficient_{13}^- + \globalcoefficient_{24}^+
\end{pmatrix}.
\end{equation}

\noindent
Since the compatibility matrix anticommutes with the permutation operator, $\permutation_d \compatibility \permutation_d = - \compatibility$, it is off-block diagonal in the eigenbasis of this operator:

\begin{equation} 
\begin{split}
\compatibility^{\text{sym}} = \mathbf{S}^{-1} \compatibility \mathbf{S} =
\begin{pmatrix}
0 & 0 & 0 & 0 \\
0 & 0 & \globalcoefficient^-_{24} & \globalcoefficient^-_{13} \\
0 & \globalcoefficient^-_{24} & 0 & 0 \\
0 & \globalcoefficient^-_{13} & 0 & 0
\end{pmatrix}, \quad
\mathbf{S} = \frac{1}{2} \begin{pmatrix}
1 & 1 & 1 & 1 \\
1 & -1 & 1 & -1 \\
1 & 1 & -1 & -1 \\
1 & -1 & -1 & 1
\end{pmatrix}, \quad
\permutation_d = \begin{pmatrix}
0 & 0 & 1 & 0 \\
0 & 0 & 0 & 1 \\
1 & 0 & 0 & 0 \\
0 & 1 & 0 & 0
\end{pmatrix}, \\
\ket{++} = \frac{1}{2} \begin{pmatrix} +1 \\ +1 \\ +1 \\ +1 \end{pmatrix}, \quad
\ket{--} = \frac{1}{2} \begin{pmatrix} +1 \\ -1 \\ +1 \\ -1 \end{pmatrix}, \quad
\ket{+-} = \frac{1}{2} \begin{pmatrix} +1 \\ +1 \\ -1 \\ -1 \end{pmatrix}, \quad
\ket{-+} = \frac{1}{2} \begin{pmatrix} +1 \\ -1 \\ -1 \\ +1 \end{pmatrix}.
\end{split}
\end{equation}

\subsection{Mapping from vertex to face amplitudes}

The vertex amplitudes, $\ket{\vertexamplitude}$, that correspond to linear isometries, $\compatibility \ket{\vertexamplitude} = \mathbf{0}$, generically induce some bending of the faces as indicated by edge compatibility for arbitrary parallelogram-based origami sheets:

\begin{equation}
\faceamplitude^{A'} - \faceamplitude^A = \vertexamplitude^{a'} \frac{\foldcoefficient_{i+2}^{a'}}{\localedge_{i+2}^{a'}} - \vertexamplitude^a \frac{\foldcoefficient_i^a}{\localedge_i^a}.
\end{equation}

\noindent
Thus, the \emph{difference} between the amplitude on a generic face and some reference face can be recursively determined by constructing a path between them. For consistency, such face amplitudes should be orthogonal to the uniform face-bending mode. For four-parallelogram origami, this procedure yields the following orthogonal basis for the linear isometries:

\begin{align}
    \ket{\vertexamplitude_+} = \ket{++}, & \quad \ket{\faceamplitude_+} = \mathbf{0}, \\
    \ket{\vertexamplitude_-} = \mathcal{N}_- \Big( \globalcoefficient_{13}^- \ket{+-} - \globalcoefficient_{24}^- \ket{-+} \Big), & \quad \ket{\faceamplitude_-} = \frac{\mathcal{N}_-}{2} \Big( \globalcoefficient_{13}^+ \globalcoefficient_{24}^- \ket{+-} + \globalcoefficient_{13}^- \globalcoefficient_{24}^+ \ket{-+} + \globalcoefficient_{13}^- \globalcoefficient_{24}^- \ket{--} \Big), \\
    \ket{\vertexamplitude_0} = \mathbf{0}, & \quad \ket{\faceamplitude_0} = \mathcal{N}_0 \ket{++},
\end{align}

\noindent
where the coefficients, $\mathcal{N}$, are normalization factors. Note that these factors cannot simultaneously impose normalization of the vertices and face but instead can be chosen to satisfy $\braket{\vertexamplitude | \vertexamplitude} + \braket{\faceamplitude | \faceamplitude} = 1$.

\begin{figure}
    \centering
    \includegraphics{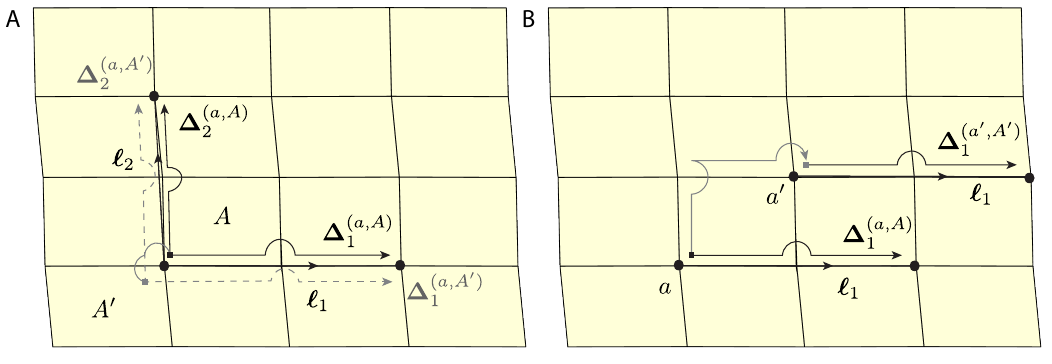}
    \caption{An illustration of the local dependence on changes to the lattice vectors. (A) Changes to the lattice vectors depend on the face the corner is defined on as indicated by the dashed gray paths in contrast to the solid black paths. Linear compatibility allows these paths to be modified, as indicated by the solid gray path, thereby relating the changes in the lattice vector on all four corners in the vicinity of a single vertex. (B) Changes in the lattice vectors depend on the vertex the corner is defined on as indicated by the two black paths. Linear compatibility implies that this difference is given by the displacement computed along the solid gray path.}
    \label{fig:stretches}
\end{figure}

\section{Lattice strain and curvature} \label{appendix:deformations}
Here, we explicitly derive the local stretches of the lattice vectors due to linear isometries and relate them to the intercellular rotations of the origami sheet.

The changes in the lattice vectors measured from a corner in the vicinity of a particular vertex are obtained by the double integration between this vertex is adjacent cells:

\begin{align}
\latticevectorchange_1^{(a,A)} & = \big( \facebend_1^A \localedgedirection_1^a - \vertexfold_2^{\permutation_h a} \localedgedirection_2^{\permutation_h a} \big) \times \localedgevector_1^{\permutation_h a} + \mathbf{f}_1(A) \times \latticevector_1, \\
\latticevectorchange_2^{(a,A)} & = \big( \facebend_4^A \localedgedirection_2^a + \vertexfold_1^{\permutation_v a} \localedgedirection_1^{\permutation_v a} \big) \times \localedgevector_2^{\permutation_v a} + \mathbf{f}_2(A) \times \latticevector_2.
\end{align}

\noindent
Here, the vertex folds, $\vertexfold_i^a$, and face torsions, $\facebend_i^A$, satisfy the compatibility conditions for linear isometries and $\mathbf{f}(A)$ accounts for any additional vertex folding due to the face that the corner is on as indicated by the solid gray paths in Fig.~\ref{fig:stretches}A. Note that such face-dependent vertex folding is always on either edge $i=2$ or $i=4$ for $\latticevectorchange_1$ and either edge $i=1$ or $i=3$ for $\latticevectorchange_2$. Furthermore, the lattice vectors can be written in terms of the local edge vectors as $\latticevector_1 = \localedgevector_1^a - \localedgevector_3^a$ and $\latticevector_2 = \localedgevector_2^a - \localedgevector_4^a$, and take this same form under any permutation.

\subsection{Lattice vector stretches}

First, consider their stretches computed by projection onto the same lattice vectors. Clearly, the face-dependent terms, $\mathbf{f}(A)$ vanish since their displacements are orthogonal to the lattice vector. Since the triple product coefficients of the torsions contain redundant edge vectors in both cases, these stretches are independent of face amplitude. In contrast, the triple product coefficients of the folds contain one nonredundant term that yields the vertex folding coefficients

\begin{align}
\latticevectorchange_1^a \cdot \latticevector_1 & = - \vertexamplitude^{\permutation_h a} \localedge^{\permutation_h a}_1 \localedge^{\permutation_h a}_3 \foldcoefficient_2^{\permutation_h a} \foldcoefficient_4^{\permutation_h a}, \\
\latticevectorchange_2^a \cdot \latticevector_2 & = \vertexamplitude^{\permutation_v a} \localedge^{\permutation_v a}_2 \localedge^{\permutation_v a}_4 \foldcoefficient_1^{\permutation_v a} \foldcoefficient_3^{\permutation_v a},
\end{align}

\noindent
where the face dependence has been dropped because it is always negligible. Since these pairings of vertex folding coefficients are invariant under all permutations, the lattice vector stretches depend locally only on the vertex amplitude:

\begin{align}
\latticevectorchange_1^a \cdot \latticevector_1 = \edgeproduct \globalcoefficient_1 \globalcoefficient_3 \braket{ a | \permutation_h | \vertexamplitude }, \\
\latticevectorchange_2^a \cdot \latticevector_2 = - \edgeproduct \globalcoefficient_2 \globalcoefficient_4 \braket{ a | \permutation_v | \vertexamplitude },
\end{align}

\noindent
where the bra $\bra{a}$ projections the ket of vertex amplitudes, $\ket{\vertexamplitude}$ onto the amplitude of vertex $a$.

\subsection{Lattice vector shears}

Now, consider the shears of the lattice vectors computed by projection onto the transverse lattice vectors. While the projections no longer eliminate the face-dependent terms, $\mathbf{f}(A) \times \latticevector$, compatibility of the linear isometries ensures that any path can be used to compute changes to the lattice vectors; in particular, the path can be constructed so that $\mathbf{f}_1(A) = \mathbf{f}_2(A) = \mathbf{f}(A)$ as illustrated by the solid gray path in Fig.~\ref{fig:stretches}A, thereby eliminating the face dependence since symmetrizing over the two lattice directions adds the terms as $\mathbf{f}(A) (\latticevector_1 \times \latticevector_2 + \latticevector_2 \times \latticevector_1)$. Careful choice of local basis for writing the lattice vectors shows that:

\begin{align}
\latticevectorchange_1^a \cdot \latticevector_2 & = \facebend_1^A \localedgedirection_1^a \times (-\localedgevector_3^a) \cdot (\localedgevector_2^a - \localedgevector_4^a) - \vertexfold_2^{\permutation_h a} \localedgedirection_2^{\permutation_h a} \times \localedgevector_1^{\permutation_h a} \cdot ( \localedgevector_2^{\permutation_h a} - \localedgevector_4^{\permutation_h a} ), \\
\latticevectorchange_2^a \cdot \latticevector_1 & = \facebend_4^A \localedgedirection_2^a \times (-\localedgevector_4^a) \cdot (\localedgevector_1^a - \localedgevector_3^a) + \vertexfold_1^{\permutation_v a} \localedgedirection_1^{\permutation_v a} \times \localedgevector_2^{\permutation_v a} \cdot ( \localedgevector_1^{\permutation_v a} - \localedgevector_3^{\permutation_v a} ),
\end{align}

\noindent
which simplifies by replacing the triple products with the associated vertex folding coefficients:

\begin{align}
\latticevectorchange_1^a \cdot \latticevector_2 & = \facebend_1^A \localedge_3^a \big( \localedge_2^a \foldcoefficient_4^a + \localedge_4^a \foldcoefficient_2^a \big) - \vertexfold_2^{\permutation_h a} \localedge_1^{\permutation_h a} \localedge_4^{\permutation_h a} \foldcoefficient_3^{\permutation_h a}, \\
\latticevectorchange_2^a \cdot \latticevector_1 & = \facebend_4^A \localedge_4^a \big( \localedge_1^a \foldcoefficient_3^a + \localedge_3^a \foldcoefficient_1^a \big) + \vertexfold_1^{\permutation_v a} \localedge_2^{\permutation_v a} \localedge_3^{\permutation_v a} \foldcoefficient_4^{\permutation_h a}.
\end{align}

\noindent
Finally, substitution of the local solutions yields:

\begin{align}
\latticevectorchange_1^a \cdot \latticevector_2 & = - \faceamplitude^A \localedge_1^a \localedge_3^a \big( \localedge_2^a \foldcoefficient_4^a + \localedge_4^a \foldcoefficient_2^a \big) + \vertexamplitude^{\permutation_h a} \localedge_3^a \localedge_4^a \foldcoefficient_1^a \foldcoefficient_2^a, \\
\latticevectorchange_2^a \cdot \latticevector_1 & = - \faceamplitude^A \localedge_2^a \localedge_4^a \big( \localedge_1^a \foldcoefficient_3^a + \localedge_3^a \foldcoefficient_1^a \big) - \vertexamplitude^{\permutation_v a} \localedge_3^a \localedge_4^a \foldcoefficient_1^a \foldcoefficient_2^a,
\end{align}

\noindent
and the symmetrized shear, $\latticevectorchange_1 \cdot \latticevector_2 + \latticevectorchange_2 \cdot \latticevector_1 = 0$, vanishes after invoking the relationship between the face and vertex amplitudes and averaging over all cells.

\subsection{Lattice curvatures}
Finally, consider the difference between two locally defined changes to the lattice vectors as indicated by the two distinct black paths shown in Fig.~\ref{fig:stretches}B. Since the linear isometries satisfy position closure, this difference is equal to a rotation of the edge between their vertices, $\globaledgevector_{(a,a')}$, by the lattice angular velocity and a rotation of the lattice vector by the angular velocity gradient between their corners:

\begin{equation}
    \latticevectorchange_{\mu}^{(a',A')} - \latticevectorchange_{\mu}^{(a,A)} = \latticerotation_{\mu} \times \globaledgevector_{(a,a')} + \big( \angularvelocity^{(a,A)} - \angularvelocity^{(a',A')} \big) \times \latticevector_{\mu}.
\end{equation}

\noindent
Projecting this difference onto the same lattice vector eliminates the second term. Moreover, since the lattice angular velocity must lie in the plane spanned by the lattice vectors, the triple product simplifies $\latticerotation_{\mu} \times \globaledgevector_{(a,a')} \cdot \latticevector_{\mu} = (\globaledgevector_{(a,a')} \cdot \normalvector) ( \latticerotation_{\mu} \times \latticevector_{\mu} \cdot \normalvector )$ Hence, the rotation of the lattice vector is characterized by the local dependence of the changes to the lattice vectors:

\begin{equation}
    \latticecurvature_{\mu \mu} \equiv \latticerotation_{\mu} \times \latticevector_{\mu} \cdot \normalvector = - \frac{ (\latticevectorchange_{\mu}^{a'} - \latticevectorchange_{\mu}^a) \cdot \latticevector_{\mu} }{\globaledgevector_{(a,a')} \cdot \normalvector},
\end{equation}

\noindent
where the corner dependence reduces to vertex dependence since the projection is onto the same lattice direction. This expression is valid for \emph{any} two vertices in the unit cell. 

The same analysis can be applied to transverse projections where the lattice vector rotations vanish after symmetrization:

\begin{equation}
    \latticecurvature_{\mu \nu} \equiv \latticerotation_{\mu} \times \latticevector_{\nu} \cdot \normalvector = - \frac{ (\latticevectorchange_{\mu}^{a'} - \latticevectorchange_{\mu}^a) \cdot \latticevector_{\nu} + (\latticevectorchange_{\nu}^{a'} - \latticevectorchange_{\nu}^a) \cdot \latticevector_{\mu} }{2 \globaledgevector_{(a,a')} \cdot \normalvector},
\end{equation}

\noindent
which satisfies equality because $\latticerotation_{\mu} \times \latticevector_{\nu} = \latticerotation_{\nu} \times \latticevector_{\mu}$ by compatibility. These off-diagonal curvatures are generated by the antisymmetric bend mode:

\begin{equation}
    \latticecurvature_{12}^{\text{asym}} = \frac{R}{4} \globalcoefficient_{13}^+ \globalcoefficient_{24}^+,
\end{equation}

\noindent
as well as the twist mode:

\begin{equation}
    \latticecurvature_{12}^{\text{twist}} = -\frac{R}{2}.
\end{equation}

\section{Lattice fundamental forms} \label{appendix:forms}
In continuous two-dimensional sheets, strain and curvature correspond to changes in the diagonal components of the first and second fundamental forms respectively. Here, we discuss this connection and derive all of the components of analogous \emph{lattice} fundamental forms for four-parallelogram origami.

\subsection{Review of fundamental forms in continuous sheets}

A continuous sheet is parameterized by coordinates on the two-dimensional surface which map to positions in the three-dimensional embedding space $\embedding = \embedding(x_1, x_2)$. The first fundamental form (metric tensor) of the sheet are the coefficients that measure arclengths on the sheet in terms of the surface coordinates. These coefficients are given by the tangent vectors, $\tangentvector_{\mu} \equiv \partial_{\mu} \embedding$, of the embedding:

\begin{equation} \label{eq:firstform}
\firstform_{\mu \nu} = \tangentvector_{\mu} \cdot \tangentvector_{\nu},
\end{equation}

\noindent
which is symmetric since the cross product is commutative. This first fundamental form becomes the identity, $\firstform_{\mu \nu} = \delta_{\mu \nu}$, when the entire sheet lies in a plane and is diagonal when the tangent vectors are orthogonal, $\tangentvector_1 \cdot \tangentvector_2 = 0$. Infinitesimal changes to this quantity, $\delta \mathbf{\firstform}$, give the strains of the sheet. The second fundamental form of the sheet are the coefficients that measure deflections of the sheet. These coefficients are given by the rotations of the tangent vectors, $\boldsymbol{\kappa}_{\mu \nu} \equiv \partial_{\mu} \tangentvector_{\nu}$, out of the plane:

\begin{equation} \label{eq:secondform}
\secondform_{\mu \nu} = \boldsymbol{\kappa}_{\mu \nu} \cdot \normalvector,
\end{equation}

\noindent
where $\normalvector = \tangentvector_1 \times \tangentvector_2$ is the local normal vector of the sheet. The invariants of the second fundamental, $\mathbf{\secondform}$, give the mean curvature, $H = \text{Tr} \mathbf{\secondform}$, and Gaussian curvature, $K = \text{Det} \mathbf{\secondform} / \text{Det} \mathbf{\firstform}$, which respectively vanish for flat and cylindrical (including flat) geometries. For initially flat sheets in particular, infinitesimal changes to the second fundamental form, $\delta \mathbf{\secondform}$, are \emph{exactly} the mean curvature, $H = \text{Tr} \delta \mathbf{\secondform}$, and the Gaussian curvature always vanishes to first-order in the deformation, $K = 0$.

\subsection{Analogous fundamental forms in discretized sheets}

A discretized origami sheet is instead parameterized by cell indices, $\mathbf{n} = (n_1, n_2)$, which map to positions in three-dimensional space via the lattice vectors, $\latticevector_{\mu}$. Since the corrugation of the origami sheet suggests it functions closer to a slab than a membrane, it is appropriate to consider the geometry of the midplane defined as the average vertex position in each cell. The corresponding tangent vectors are the lattice vectors which do not rotate between cells so that the ground states have first and second lattice fundamental forms:

\begin{align}
    \firstform_{\mu \nu} & = \latticevector_{\mu} \cdot \latticevector_{\nu}, \label{eq:latticefirstform} \\
    \secondform_{\mu \nu} & = 0. \label{eq:latticesecondform}
\end{align}

Recall that the lattice vectors are generically non-orthotropic, $\latticevector_{\mu} \cdot \latticevector_{\nu} \neq 0$, so that the off-diagonal components of the first lattice fundamental form in Eq.~\ref{eq:latticefirstform} are generically non-vanishing; however, since this dot product is an invariant as indicated in Eq.~\ref{eq:orthotropic} it cannot change for rigid deformations which preserve the edge lengths and sector angles. Moreover, the first fundamental form can be diagonalized by performing a change of basis from the lattice vectors, $\latticevector_{\mu}$, to a pair of orthogonal basis vectors, $\latticevector'_{\mu}$. Without loss of generality let the first basis vector be identical to the first lattice vector, $\latticevector'_1 \equiv \latticevector_1$. The second basis vector is then obtained by Gram-Schmidt orthogonalization: $\latticevector'_2 \equiv \latticevector_2 - (\latticevector_2 \cdot \hat{\latticevector}'_1) \hat{\latticevector}'_1$. The transformation between surface coordinates can then be obtained by application of the chain rule to the line elements

\begin{equation}
    \Delta s^2 = \firstform_{\mu \nu} \frac{\partial n_{\mu}}{\partial n_{\alpha}'} \frac{\partial n_{\nu}}{\partial n_{\beta}'} \Delta n'_{\alpha} \Delta n'_{\beta} = \firstform_{\alpha \beta}' \Delta n'_{\alpha} \Delta n'_{\beta},
\end{equation}

\noindent
and inverting the partial derivatives. This transformation can be applied to the strain and curvature of the lattice along orthogonal directions with the caveat that the Poisson's ratios are no longer equal and opposite.

\subsection{Screw-periodic origami}

More generic crease patterns have screw-periodic (cylindrical) ground states~\cite{mcinerney2020hidden} for which the lattice vectors rotate between cells via the lattice rotations, $\cellrotation$, satisfying

\begin{align}
    \cellrotation_1 \cellrotation_2 & = \cellrotation_2 \cellrotation_1, \label{eq:orientationcompatibility} \\
    \latticevector_1 + \cellrotation_1 \latticevector_2 & = \latticevector_2 + \cellrotation_2 \latticevector_1. \label{eq:spatialcompatibility}
\end{align}

\noindent
The computation of the second lattice fundamental form would follow as:

\begin{equation}
    \secondform_{\mu \nu} = \cellrotation_{\mu} \latticevector_{\nu} \cdot \normalvector,
\end{equation}

\noindent
where $\normalvector = \widehat{\latticevector_1 \times \latticevector_2}$ so that this quantity is automatically symmetric via Eq.~(\ref{eq:spatialcompatibility}). In this case, the zeroth-order second fundamental form \emph{does not} correctly capture the curvature of the cylinder that this origami sheet discretizes; hence, this formalism must be augmented for more generic periodicities.

\end{document}